\documentclass[%
 reprint,
 amsmath,amssymb,
 aps,
]{revtex4-1}

\pdfoutput=1

\usepackage{graphicx}
\usepackage{dcolumn}
\usepackage{bm}

\usepackage{subfigure}

\begin{document}

\preprint{APS/123-QED}

\preprint{APS/123-QED}

\title{Boolean gates on actin filaments}

\author{Stefano Siccardi}
\email{ssiccardi@2ssas.it}
\affiliation{The Unconventional Computing Centre, University of the West of England, Bristol, UK}
 \author{Jack A. Tuszynski}
 \email{jack.tuszynski@gmail.com}
 \affiliation{
 Department of Oncology, University of Alberta, Edmonton, Alberta, Canada
}%
\author{Andrew Adamatzky}%
 \email{andrew.adamatzky@uwe.ac.uk}
\affiliation{%
 The Unconventional Computing Centre, University of the West of England, Bristol, UK\\
}%

\date{\today}

\begin{abstract}
\noindent
Actin is a globular protein which forms long polar filaments in the eukaryotic cytoskeleton. Actin networks play a key role in cell mechanics and cell motility. They have also been implicated in information transmission and processing, memory and learning in neuronal cells. The acting filaments have been shown to support propagation of voltage pulses.  Here we apply a coupled nonlinear transmission line model of actin filaments to study interactions between voltage pulses. By assigning a logical {\sc Truth} to the presence of a voltage pulses in a given location of the actin filament, and {\sc False} to the pulse's absence we represent digital information transmission along these filaments. When two pulses, representing Boolean values of input variables, interact, then they can facilitate or inhibit further propagation of each other.  We explore this phenomenon to  construct Boolean logical gates and a one-bit half-adder with interacting voltage pulses. 
We discuss implications of these findings on cellular process and technological applications. 
\end{abstract}

\maketitle

Actin  is a globular structural protein; one of the most highly conserved eukaryotic protein, ranging from unicellular organisms to plants and animals.  Actin plays a key role in cell motility forming actin filaments (microfilaments), which in turn generate parallel bundles that the cell utilizes for contractility, in cell division for the creation of the cleavage furrow and in cell motility. Actin has also been correlated with nervous system activity and learning: actin cytoskeleton affects synaptic properties leading to learning, actin and its regulatory proteins are involved in various stages of memory~\cite{tuszynski1998dielectric,priel2006dendritic, debanne2004information, priel2010neural, jaeken2007new}.  The actin filament networks are key players in modulation of synaptic terminals due to their interactions with ion channels~\cite{cantiello1997role} and filtration of noise in synapses~\cite{dillon2005actin, cingolani2008actin}.   By modulating dendritic ion channel activity actin filaments govern neural information processing and facilitate computational abilities of dendritic trees via facilitation of ionic condensation and ion cloud propagation~\cite{priel2005electrodynamic}.

Previously we proposed a model of  actin filaments as two chains of one-dimensional binary-state semi-totalistic automaton arrays to signalling events, and discovered local activity rules that supports travelling or stationary localizations~\cite{adamatzky2014actin}.  This finite state machine model has been further extended to a quantum cellular automata (QCA) model in \cite{siccardi2015actin}. We have shown that quantum actin automata can perform basic operations of Boolean logic, and implemented a binary adder~\cite{siccardi2015actin}, and three valued logic operations~\cite{cadamatzky}. These models were implemented in general algorithmic terms, without describing any
specific physical mechanisms that could be used to implement cell excitations and interactions. They also did not employ interactions between propagating localisations in a spirit of collision-based computing~\cite{adamatzky2000collision}. 

We aim to rectify these omissions and present a model of actin in terms of RLC (resistance, capacitance, inductance) non-linear electrical transmission wires that can implement logical gates via interacting voltage impulses. These models take inspiration from the idea of a cellular automaton,
but do not fit exactly in its definition, as cells do not possess discrete states, there are no intrinsic 
time steps, and no rules are defined for state transitions. However, the states can be digitised by defining
a suitable threshold, a time step can be defined by convenience and transitions are governed by Kirchhoff's circuit laws so that a deterministic evolution takes place even in absence of explicit transition rules.

Our starting point is the usual model of electrical wires as sequences of circuits composed by resistors, capacitors and inductors. Our model is based on Tuszynski et al.~\cite{tuszynski2004ionic} model of actin monomers in terms of electrical components. This latter model was developed in order to explain experimental observations of ionic conductivity along actin filaments~\cite{lin1993novel}. This model exhibits, in a continuous limit, solitons (standing waves and travelling impulses). 

The aim of the present work is to use the same equations as in.~\cite{tuszynski2004ionic} but without invoking the continuous limit approximation
and to study the behaviour of several types of solutions for tens of monomers. We will show how an excitation moves along the actin filament and how it collides with another excitation coming from elsewhere; we will show that these collisions can be used to implement logical operations. This could provide a physical description of signal propagation and processing in the cellular milieu and also possibly in hybrid bio-nano technological applications.

\section{Actin filaments as nonlinear RLC transmission lines.}

\begin{figure}[h]
		\includegraphics[width=1\linewidth]{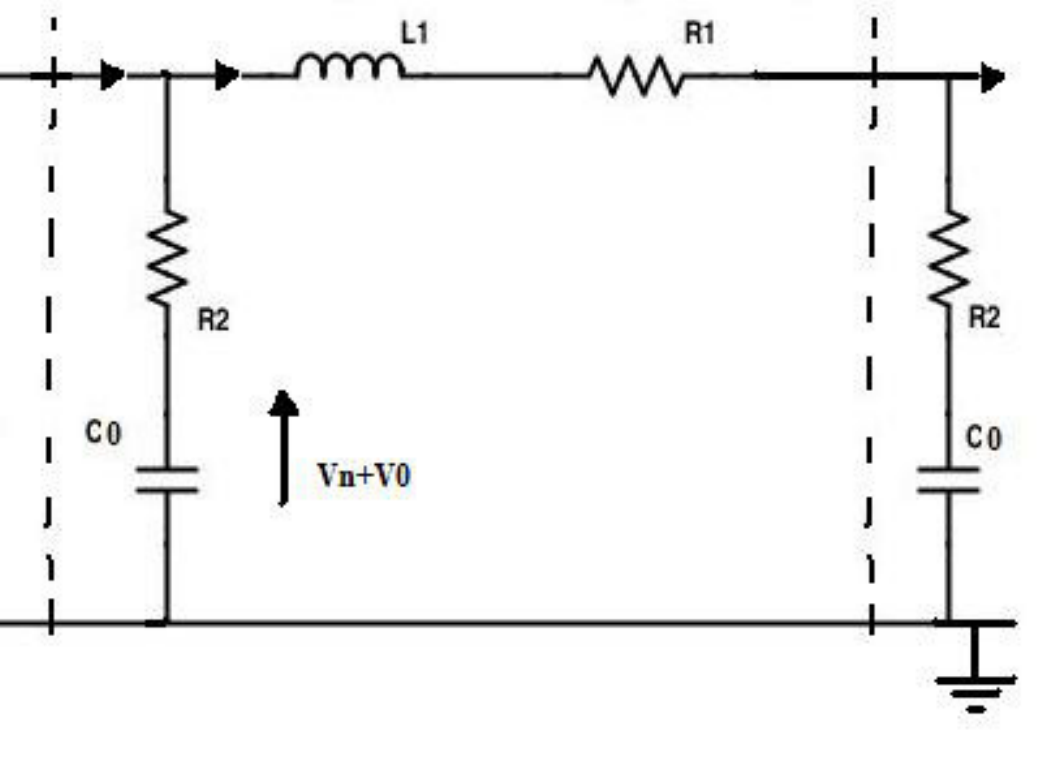}
	\caption{A circuit diagram for the $n$-th unit of an actin filament (from \cite{tuszynski2004ionic}).}
	\label{modello}
\end{figure}

In this section we recall the basic model that will be used throughout the paper; more details
can be found in the original paper \cite{tuszynski2004ionic}.

Referring to Fig.~\ref{modello}, where an actin monomer unit in a filament is delimited by the dotted lines,
 we assume  that capacitors are nonlinear (see discussion in \cite{ma1999weakly}, \cite{wang1999nonlinear}):

\begin{equation}
Q_{n} \,=\, C_{0} (V_{n} \,-\, bV_{n}^{2})
\label{equcapacitor}
\end{equation}

The main equation is:

$
L C_{0} \frac{d^{2}}{dt^{2}} (V_{n} - b V_{n}^{2}) \,=
$

$
=\, V_{n+1} + V_{n-1} - 2V_{n} \,-\, R_{1}C_{0} \frac{d}{dt} (V_{n} - b V_{n}^{2})
$

$
-\, R_{2}C_{0} \{ 2 \frac{d}{dt} (V_{n} - b V_{n}^{2}) - \frac{d}{dt} (V_{n+1} - b V_{n+1}^{2})
$

\begin{equation}
-\, \frac{d}{dt} (V_{n-1} - b V_{n-1}^{2}) \}
\label{secondorder}
\end{equation}

The above is the basic equation, that will be used to compute the cells' behaviour.

In \cite{tuszynski2004ionic} formulas are derived to obtain values for the relevant parameters. The Bjerrum length is defined as follows:
\begin{equation}
\lambda_{B} = \frac{e^2}{4 \pi \epsilon \epsilon_0 k_B T}
\label{bjerrum}
\end{equation}
where $e$ is the electrical charge, $\epsilon_0$ the permittivity of the vacuum, $\epsilon$ the dielectric
constant of the solution where the actin molecule is placed, estimated similar to $\epsilon_{water} \approx 80$,
$k_B$ Boltzmann's constant and $T$ the absolute temperature, that is taken $\approx 293 K$.
The Bjerrum length is the distance beyond which thermal fluctuations are stronger that electrostatic
interactions between charges. The parameter used in  \cite{tuszynski2004ionic}  is $\lambda_B = 7.1 \times 10^{-10} m$.

The formula for the capacitance is as follows:
\begin{equation}
C_0 = \frac{2 \pi \epsilon \epsilon_0 l}{ln ( \frac{r_{actin} + \lambda_B}{r_{actin}} )}
\label{capacitance}
\end{equation}
where $l$ is the length of a monomer, taken $\approx 5.4 nm$; and $r_{actin} \approx 2.5 nm$.
With the parameters given, the capacitance per monomer is estimated $C_0 \approx 96 \times 10^{-6} pF$.

For the inductance we have:
\begin{equation}
L = \frac{\mu N^2 \pi (r_{actin} + \lambda_B)^2}{l}
\label{inductance}
\end{equation}
where $\mu$ is the magnetic permeability and $N$ is the number of turns of the coil, that is the number of
windings of the distribution of ions around the filament. It is approximated by counting how many ions can
be lined up along the length of a monomer as $N=l/r_h$, and it is supposed that the size of a typical ion
is  $r_h \approx 3.6 \times 10^{-10} m$.
With these assumptions we have for a monomer $L \approx 1.7 pH$.

For the resistance we have
\begin{equation}
R = \frac{\rho \, ln ((r_{actin} + \lambda_B)/r_{actin})}{2 \pi l}
\label{resistance}
\end{equation}
where resistivity $\rho$ is approximately given by:
\begin{equation}
\rho = \frac{1}{\Lambda_{0}^{K^{+}} c_{K^{+}}+\Lambda_{0}^{Na^{+}} c_{Na^{+}}}
\label{resistivity}
\end{equation}
where $c_{K^{+}}$ and $c_{Na^{+}}$ are the concentrations of sodium and potassium ions, considered
respectively 0.15~M and 0.02~M; $\Lambda_{0}^{K^{+}} \approx 7.4 (\Omega m)^{-1} M^{-1}$ and 
$\Lambda_{0}^{Na^{+}}\approx 5.0 (\Omega m)^{-1} M^{-1}$ depends only on the type of salts.
Accordingly, $R_1 \approx 6.11 M\Omega$; $R_2$ is taken $= R_{1}/7 \approx 0.9 M\Omega$ following 
\cite{tuszynski2004ionic}.

We note that resistance, inductance and capacity values could be tuned by changing the temperature and the ion concentrations. Further sections deal with  solutions of (\ref{secondorder})
and their interactions.

\section{Evolution of pulses}

We recap the parameters that will be used in almost all computations:
\begin{equation}
\begin{matrix}
L=1.7 {\text pH},\; C_{0}=96\cdot 10^{-6} {\text pF}, \\ R_{1}=6.11\cdot 10^6 \Omega, \; R_{2}=0.9\cdot 10^6 \Omega
\end{matrix}
\label{valorig}
\end{equation}
With these parameters we measure time in nanoseconds.
Let us consider the case when an initial voltage is set at one end of the filament for some time as, for example, could take place when an actin filament is in close proximity of an ion channel. We use for voltage the form:
\begin{equation}
V_0 = \frac{1}{2} - \frac{e^{t-t_0}-e^{-t+t_0}}{2(e^{t-t_0}+e^{-t+t_0})}
\label{inistep}
\end{equation}
In all the following computations we consider $t_0=3$~ns.
\begin{figure}[!tbp]
\centering
\subfigure[]{\includegraphics[width=1\linewidth]{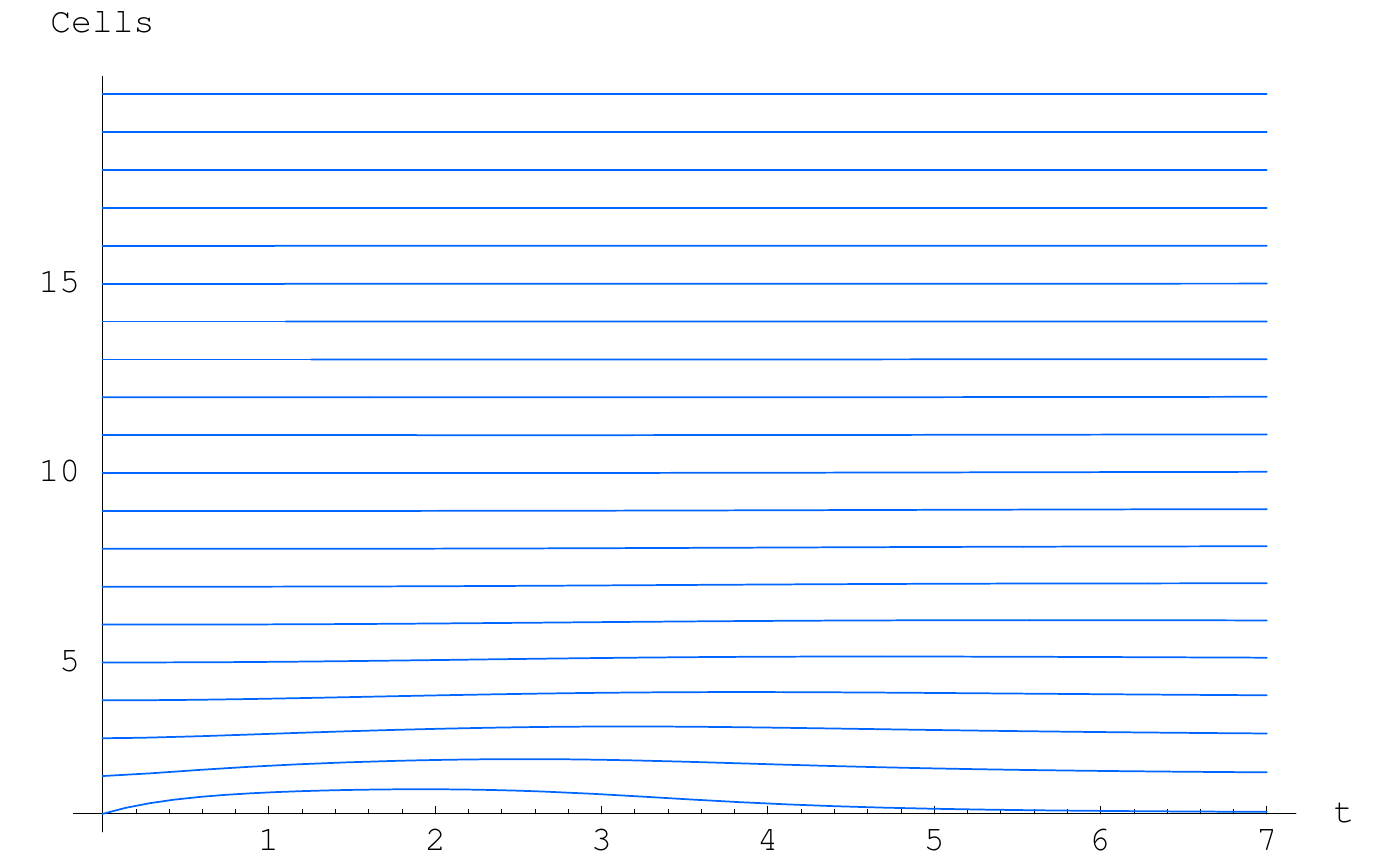}}
\subfigure[]{\includegraphics[width=1\linewidth]{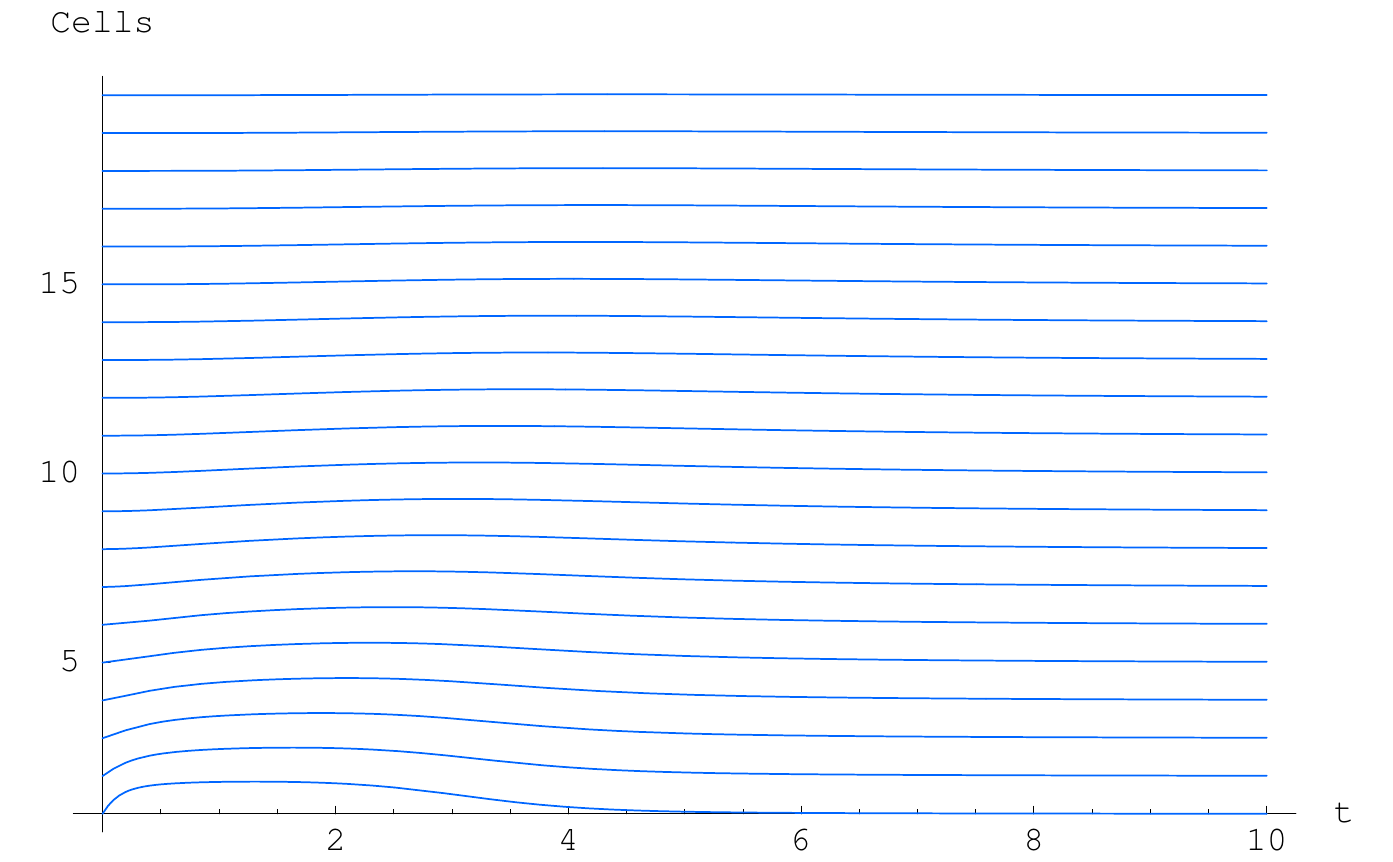}}
\caption{An input pulse travelling along actin filament: 
(a)~parameters (\ref{valorig}), (b)~the same parameters but $R_1=61.1 \, 10^5 \Omega$. The pictures represent the cell evolutions, each cell voltage is represented by a diagram with time on the horizontal axis and voltage $V$ on the vertical one. The diagrams are tiled vertically with cell 1 at the
bottom and cell 20 at the top, with a unitary displacement.}
	\label{travel_tuz}
\end{figure}

We find that pulses actually travel, but after a few nanoseconds they are damped (Fig.~\ref{travel_tuz}a).
From the diagram we see that the 12th unit is reached in about 4~ns, so that the speed is approximately $5.4$~nm $\times 12 / 4$~ns, that is about 16~m/sec. That is consistent with the estimates in \cite{tuszynski2004ionic}.
If we decrease the resistance $R_{1}$ by  an order of magnitude, the speed doubles. Then we can obtain a much longer lasting pulse (Fig.~\ref{travel_tuz}b).

\begin{figure}[!tbp]
\centering
\subfigure[]{\includegraphics[width=1\linewidth]{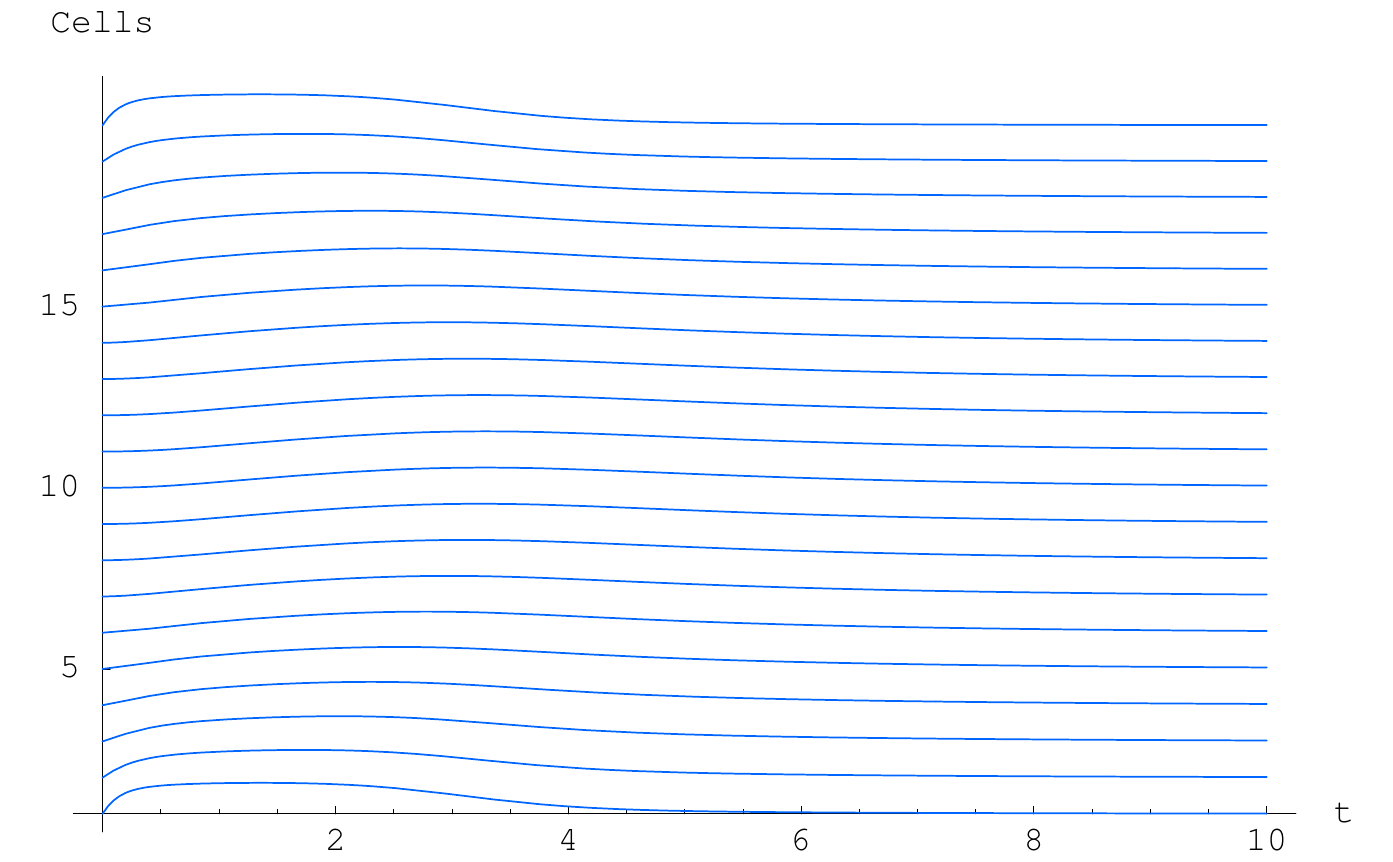}}
\subfigure[]{\includegraphics[width=1\linewidth]{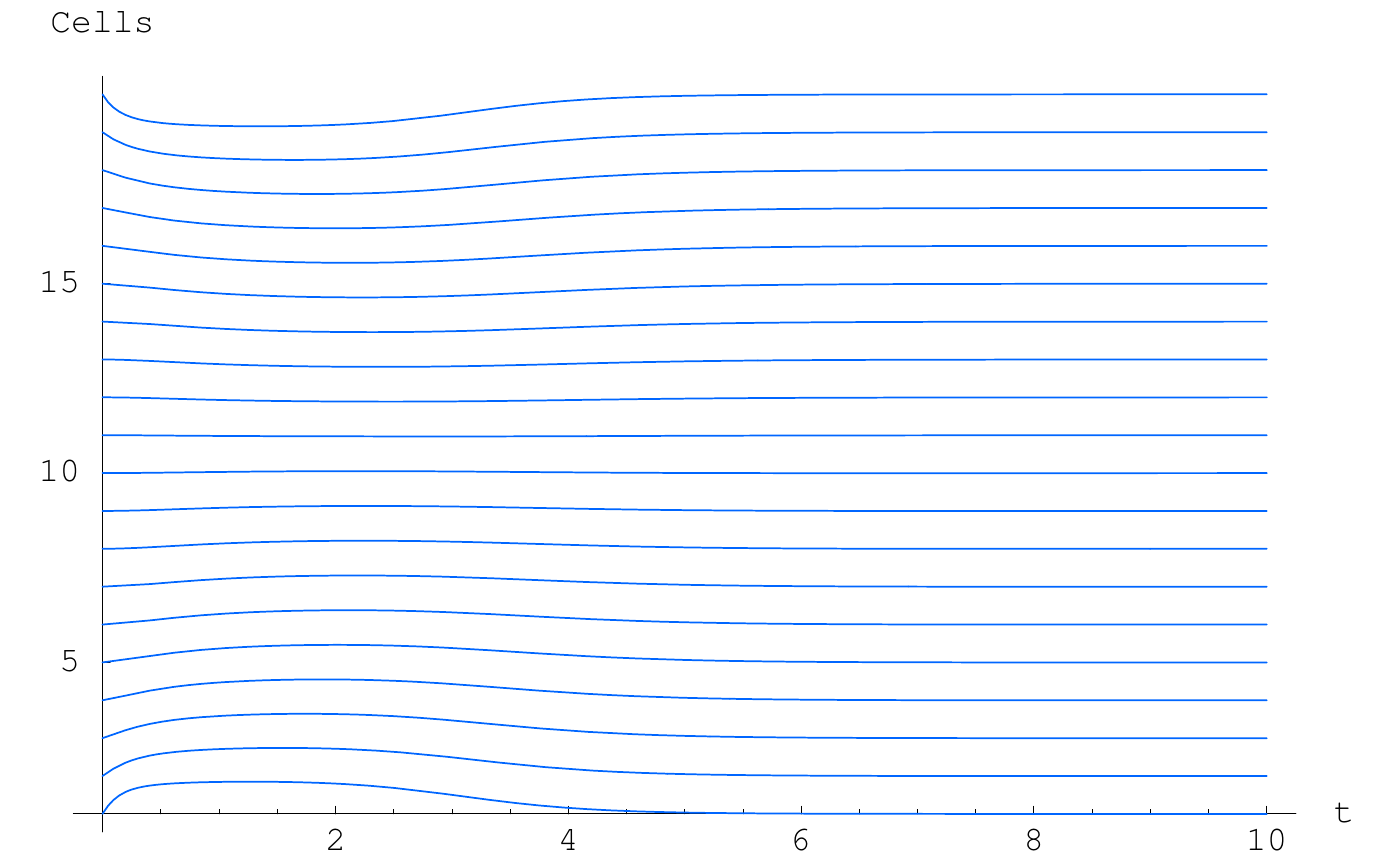}}
\caption{Two input pulses travelling along actin monomers from the opposite ends, with parameters (\ref{valorig}), but $R_1=61.1 \, 10^5 \Omega$. (a)~Identical pulses. (b)~Opposite voltage.}
	\label{travel_tuz3}
\end{figure}

Figure~\ref{travel_tuz3}a shows the case where two identical impulses are applied at the ends of the actin filament. 
In Fig.~\ref{travel_tuz3}b we see interaction of two pulses of opposite sign applied to ends of the filament.

 In the next two sections we consider two types of inputs: forced and unforced. An unforced input is an impulse applied just once at input sites of actin-based gate. A forced input is an impulse applied continuously during evolution of the gate.

\section{Gates with unforced pulses}

Let us consider interactions and possible gates implemented using only initial conditions of cells, without any forced pulses applied. They are based on the assumption  that we can initialize any cells with a pre-determined
electrical potential value and to reliably measure output potentials at given units of actin filament.

Using parameters (\ref{valorig}) and $b=0.1$, and  initial condition $V=V_0 \neq 0$, we look at cells having voltages greater than a fixed percentage of initial potential $V_0$. We find that some logical operations are possible to implement, if we consider signals above $\approx 0.1 V_0$ as output, to read outputs of the logical gates.

\begin{figure}[!tbp]
	\centering
		\subfigure[]{\includegraphics[width=0.49\linewidth]{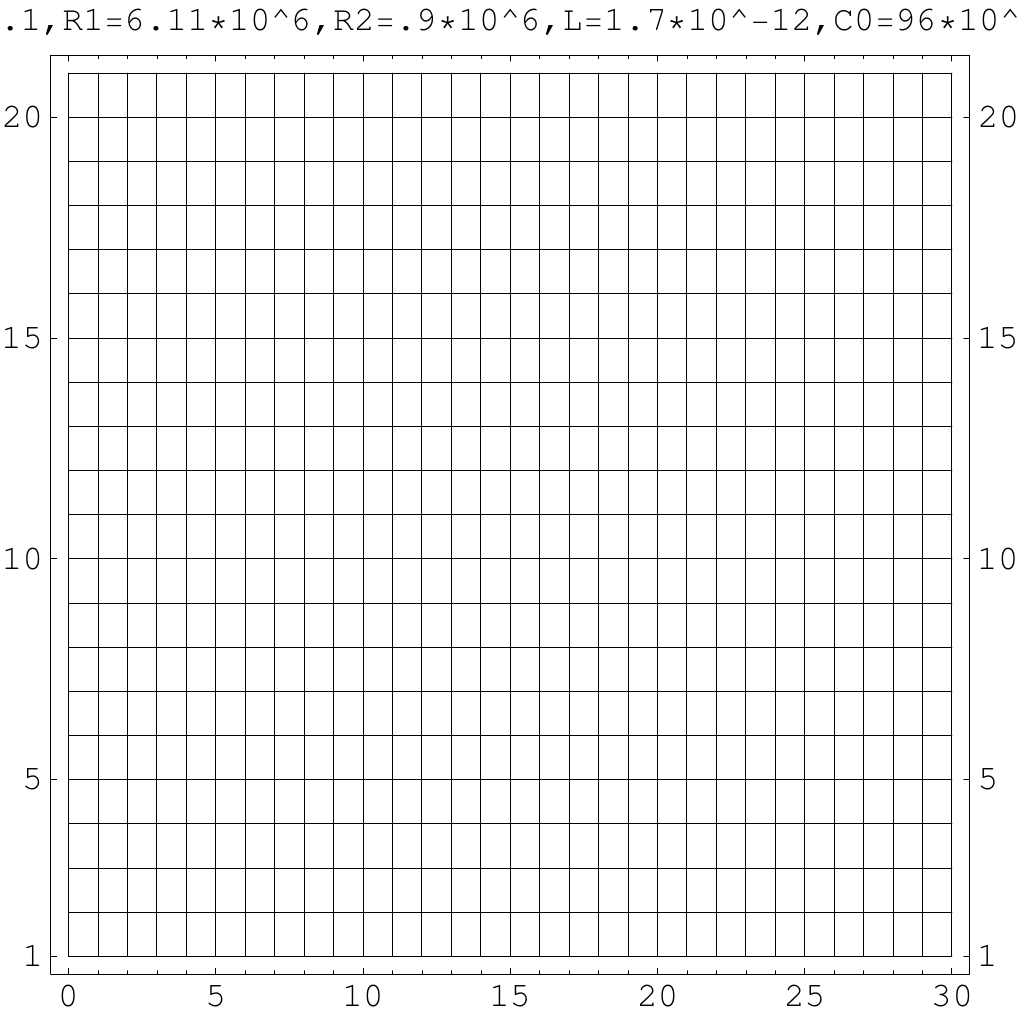}}
		\subfigure[]{\includegraphics[width=0.49\linewidth]{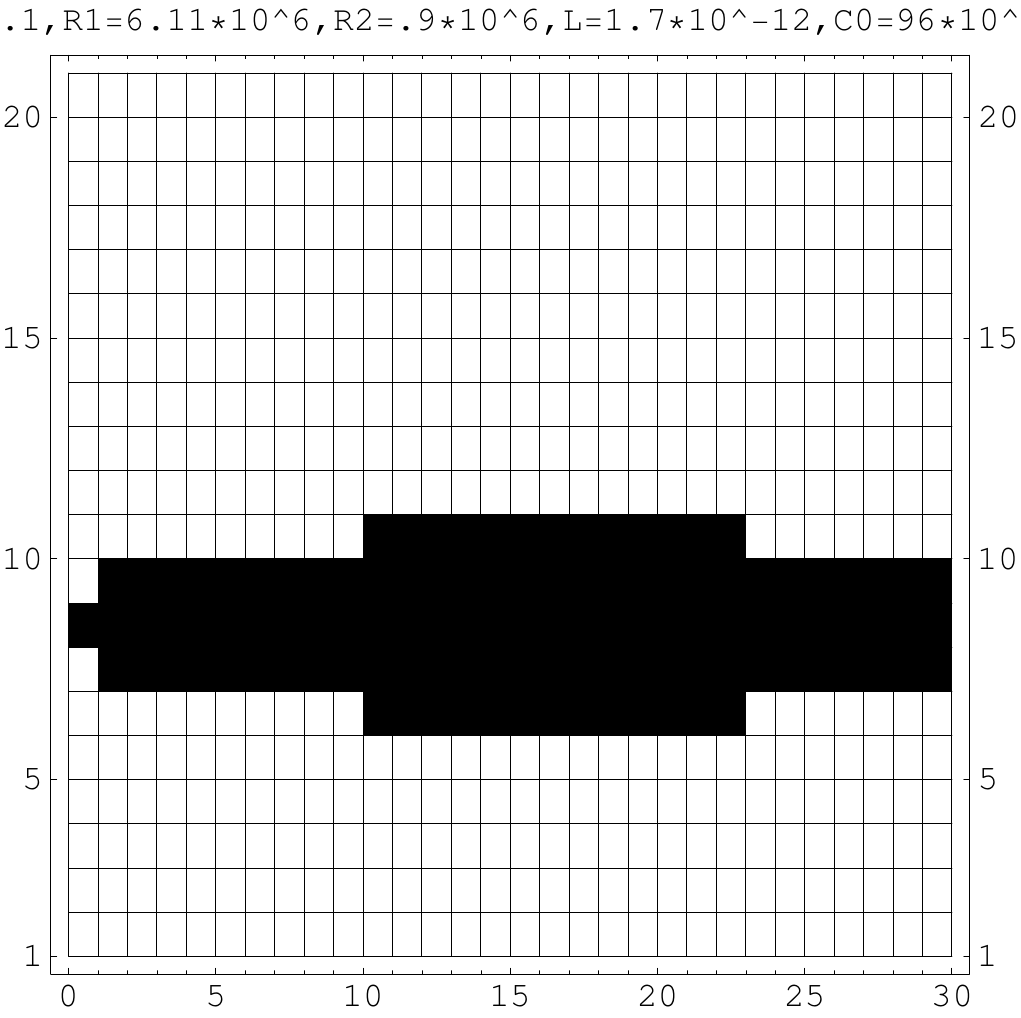}}
		\subfigure[]{\includegraphics[width=0.49\linewidth]{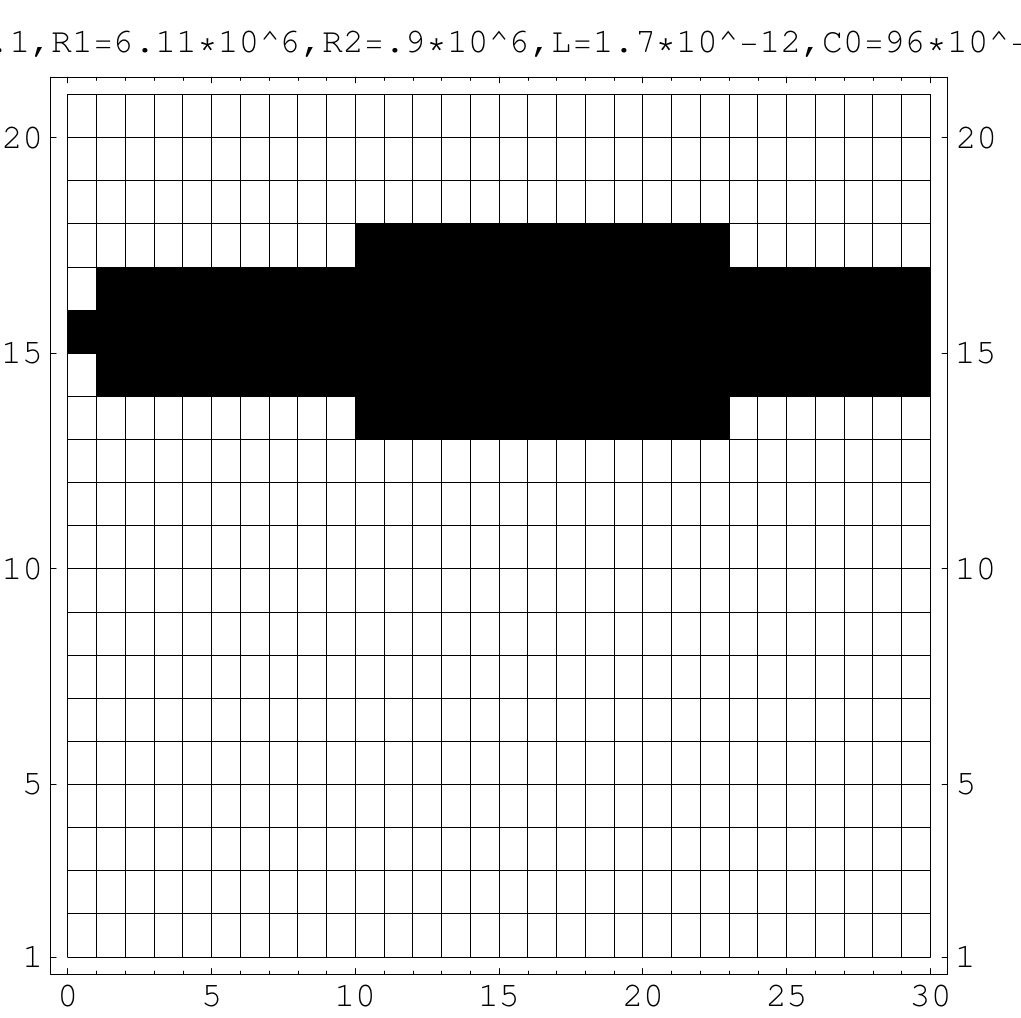}}
		\subfigure[]{\includegraphics[width=0.49\linewidth]{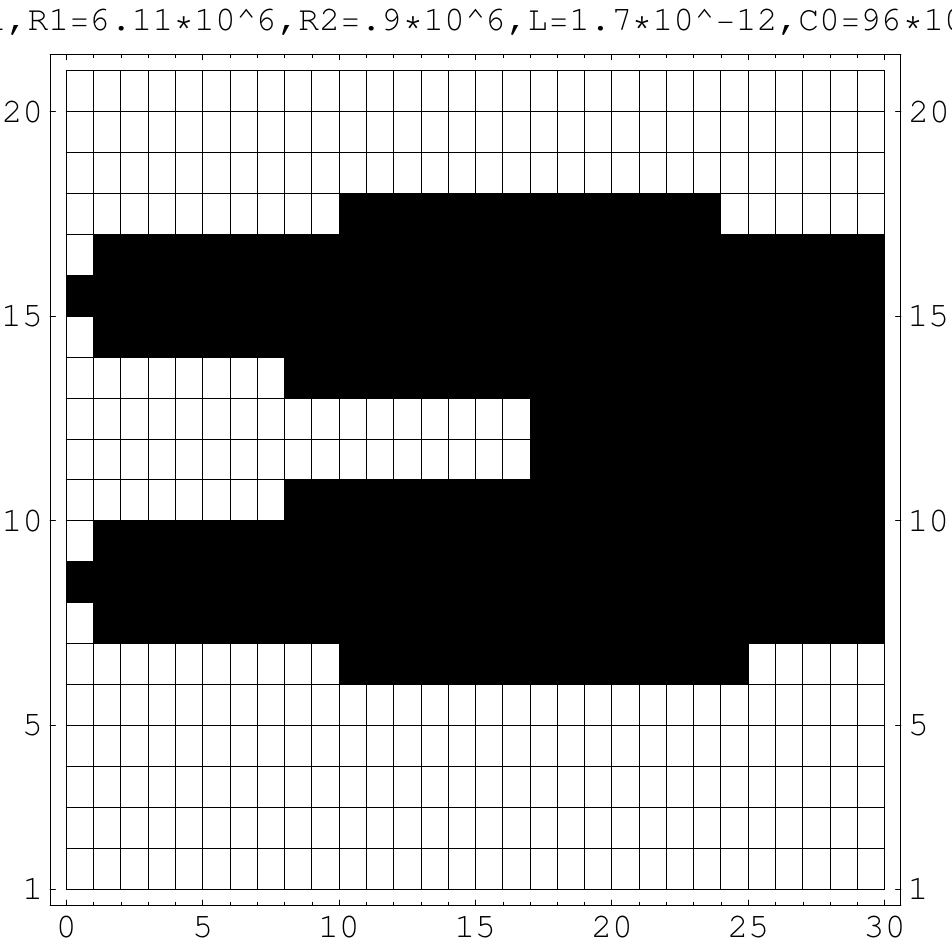}}
		\caption{Unforced pulses, {\sc and} operation. Inputs are (a)~00, (b)~10, (c)~01 and (d)~11 in cells 8 and 15; output in cells 11 and 12. Cells excited above $0.1 V_0$ are black; time evolves along the horizontal axis and cells are aligned along the vertical axis.}
		\label{and_tuz}
\end{figure}

\begin{figure}[!tbp]
	\centering
	\subfigure[]{\includegraphics[width=0.49\linewidth]{and00_tuz}}
	\subfigure[]{\includegraphics[width=0.49\linewidth]{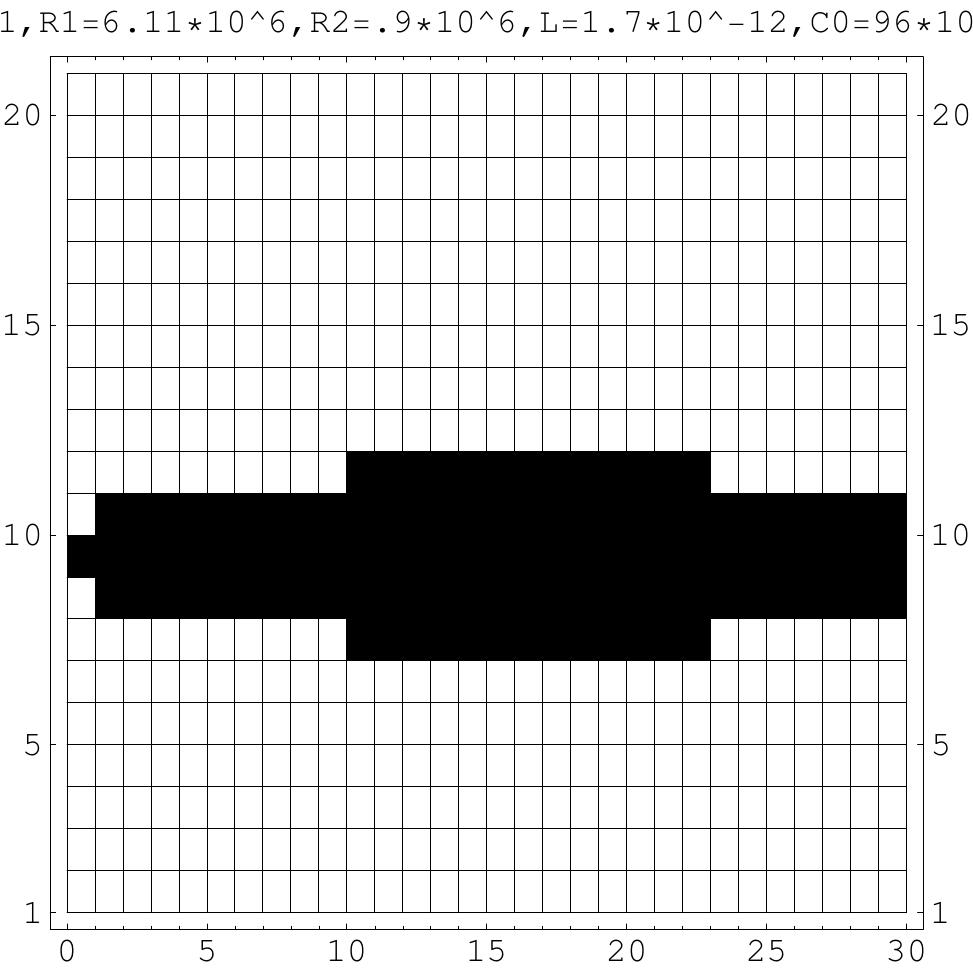}}
	\subfigure[]{\includegraphics[width=0.49\linewidth]{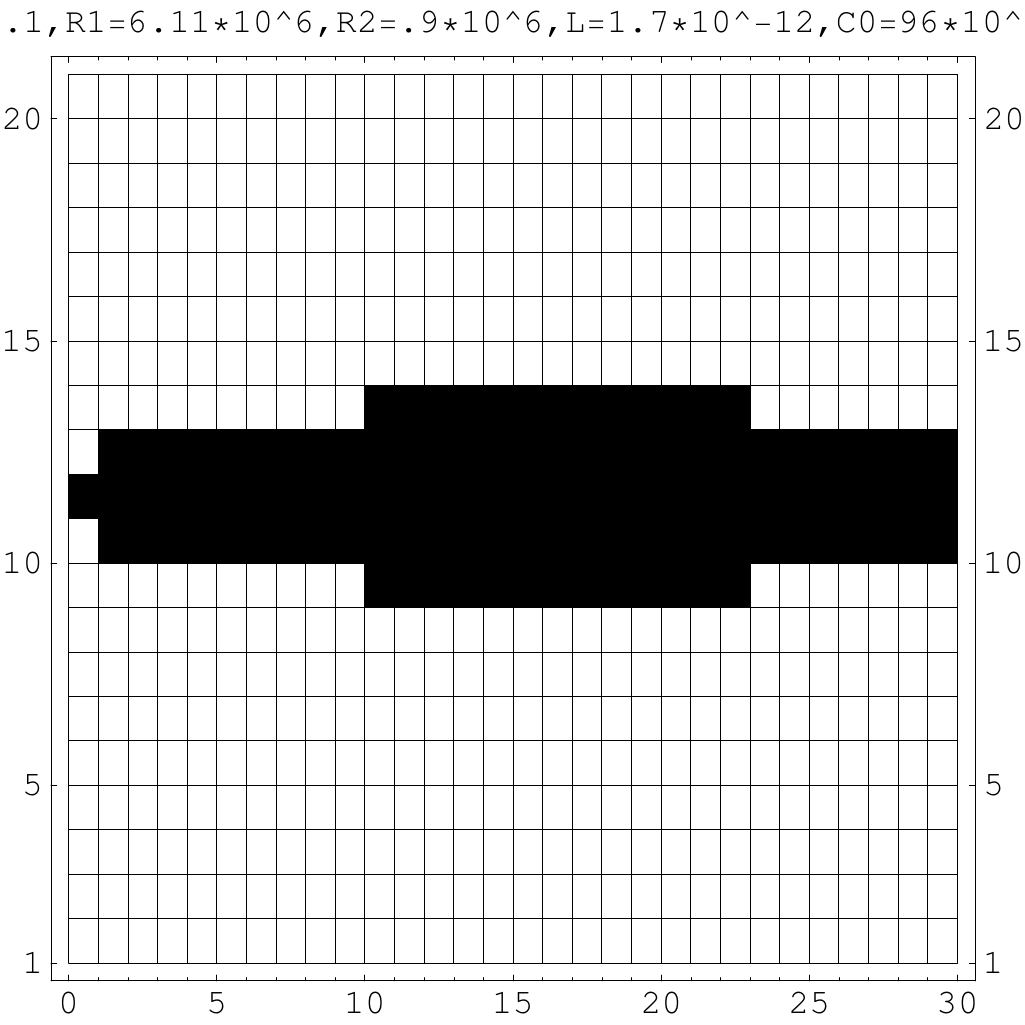}}
	\subfigure[]{\includegraphics[width=0.49\linewidth]{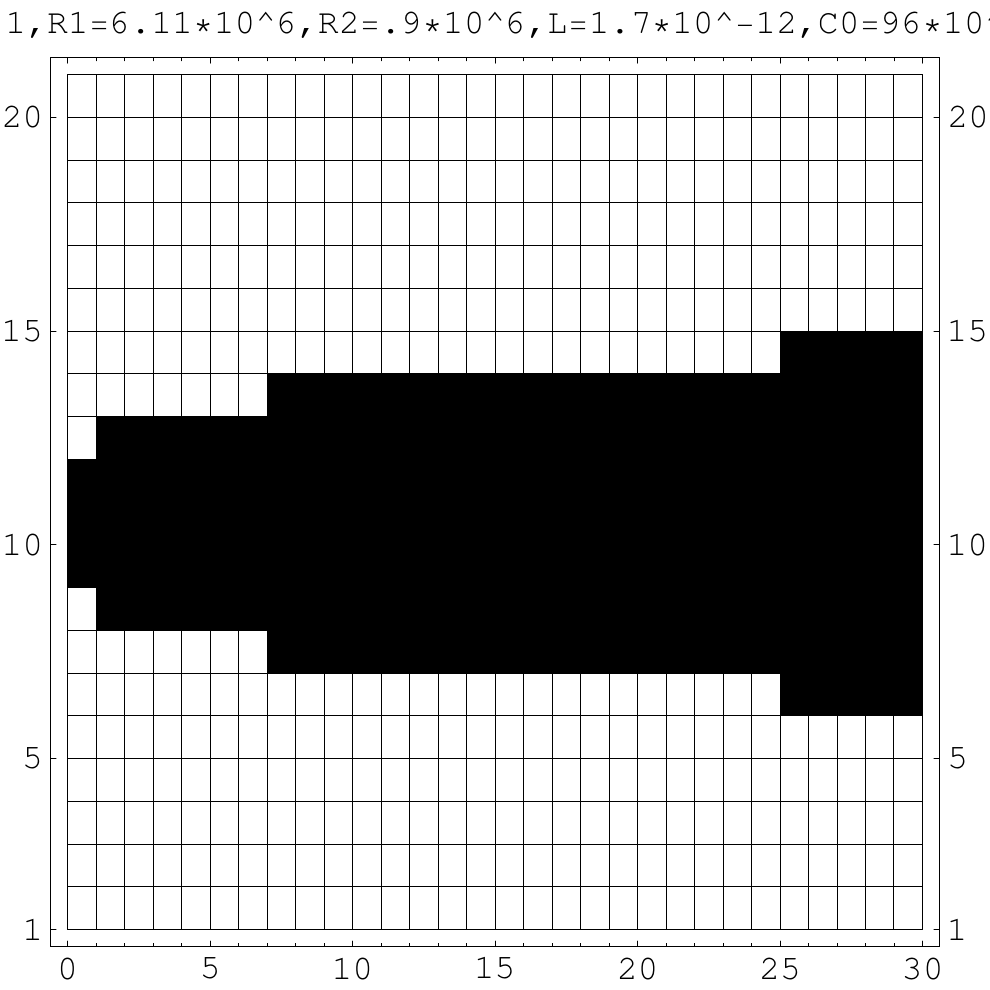}}
		\caption{Unforced pulses, {\sc or} operation. Inputs are
		(a)~00, (b)~10, (c)~01 and (d)~11 in cells 9 and 11; output in cell 10. Cells excited above $0.1 V_0$ are black; time evolves along the horizontal axis and cells are aligned along the vertical axis.}
		\label{or_tuz}
\end{figure}

Figure~\ref{and_tuz} shows the {\sc and} operation obtained with two input cells at positions 8 and 15.
The output is  in cells 11 and 12, that are excited at the end of the evolution if both the input cells were excited initially.
Figure~\ref{or_tuz} shows the {\sc or} operation obtained with two input cells at positions 9 and 11. 
The output is in cell 10 that is excited at the end of the evolution  if  one of the input cells or both of them were excited initially.

\begin{figure}[!tbp]
	\centering
		\subfigure[]{\includegraphics[width=0.49\linewidth]{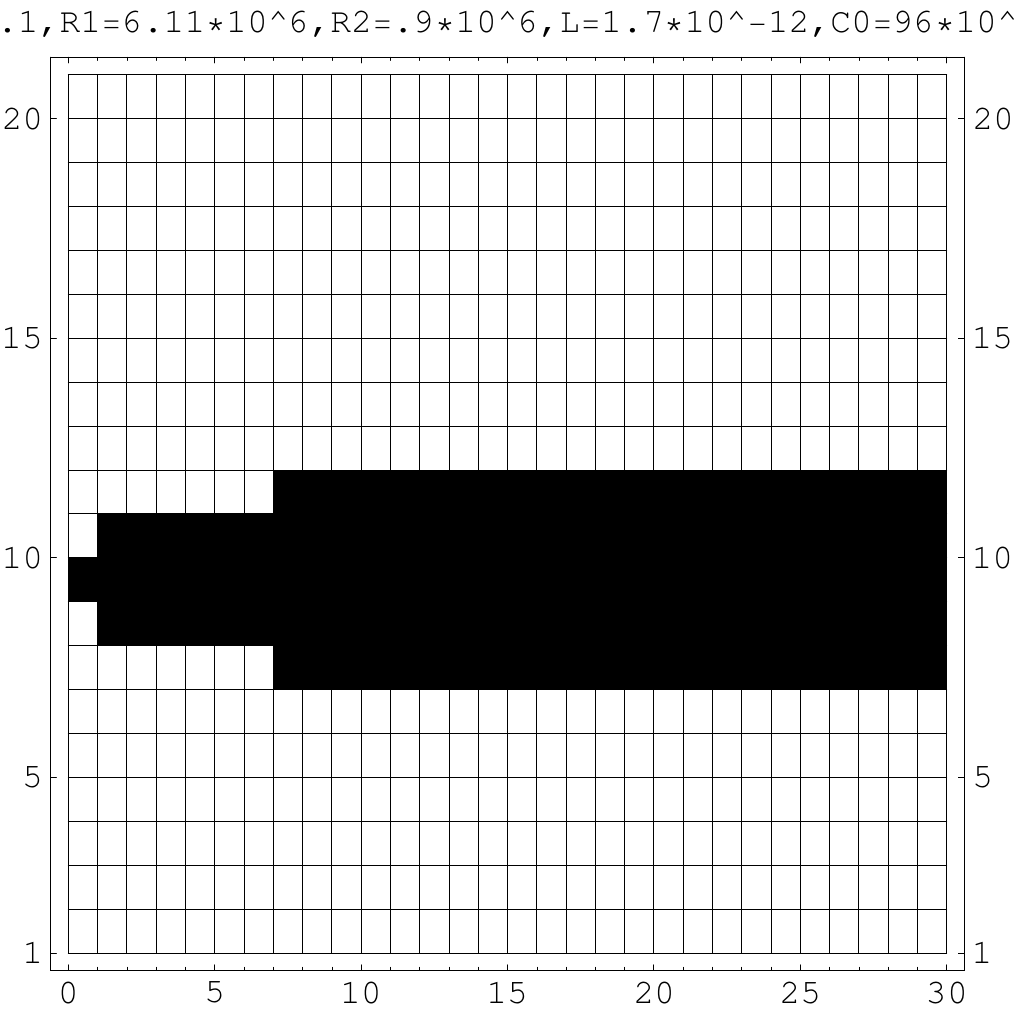}}
		\subfigure[]{\includegraphics[width=0.49\linewidth]{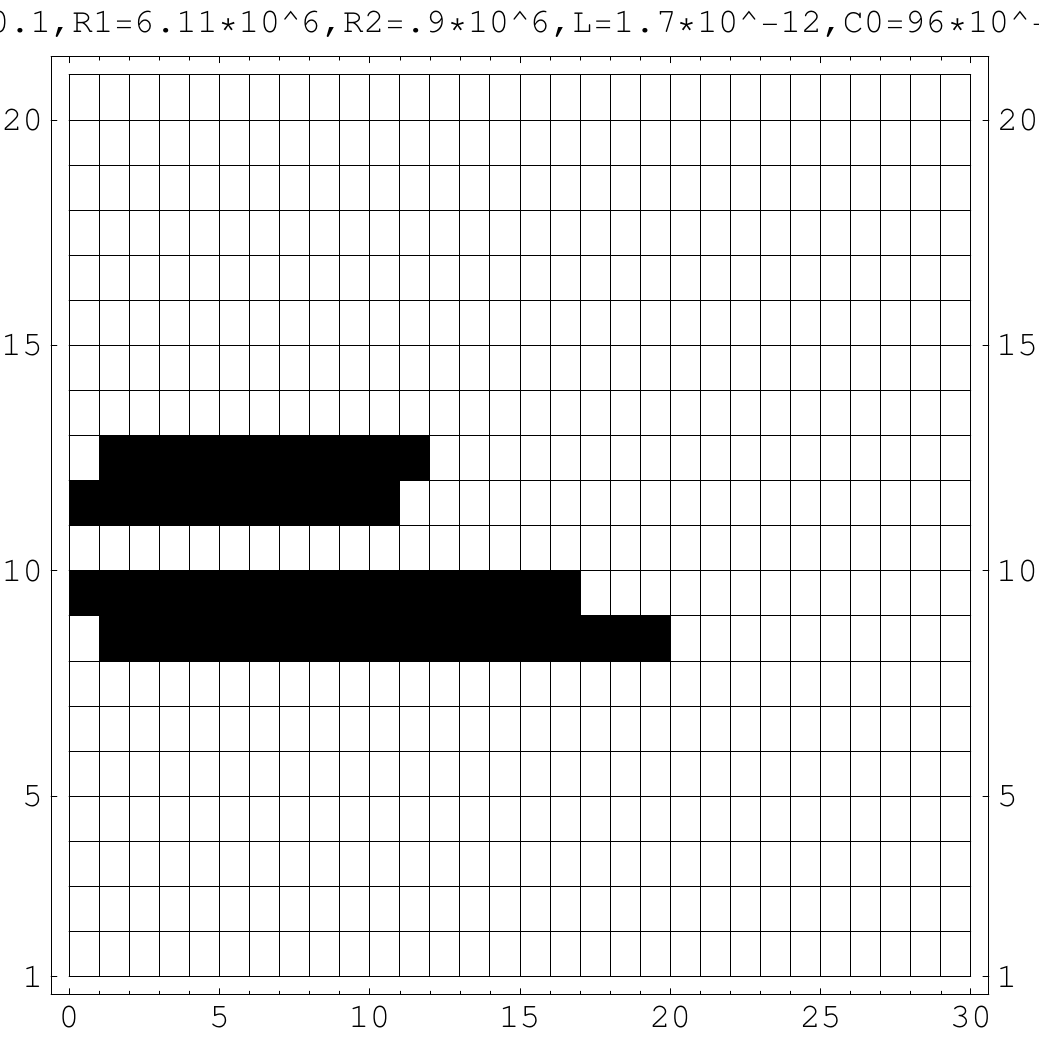}}
		\caption{Unforced pulses, {\sc not} operation.  Input is (a)~0 and (b)~1 in cell 11; output in cell 10. Cells excited above $0.1 V_0$ are black; time evolves along the horizontal axis and cells are aligned along the vertical axis.}
		\label{not_tuz}
\end{figure}

To obtain a {\sc not} gate, we initialize an auxiliary cell at position 9 at $-V_0$ and place the input cell at position 11; the output is found in cell 10 as shown in Fig.~\ref{not_tuz}.

\begin{figure}[!tbp]
	\centering
		\subfigure[]{\includegraphics[width=0.49\linewidth]{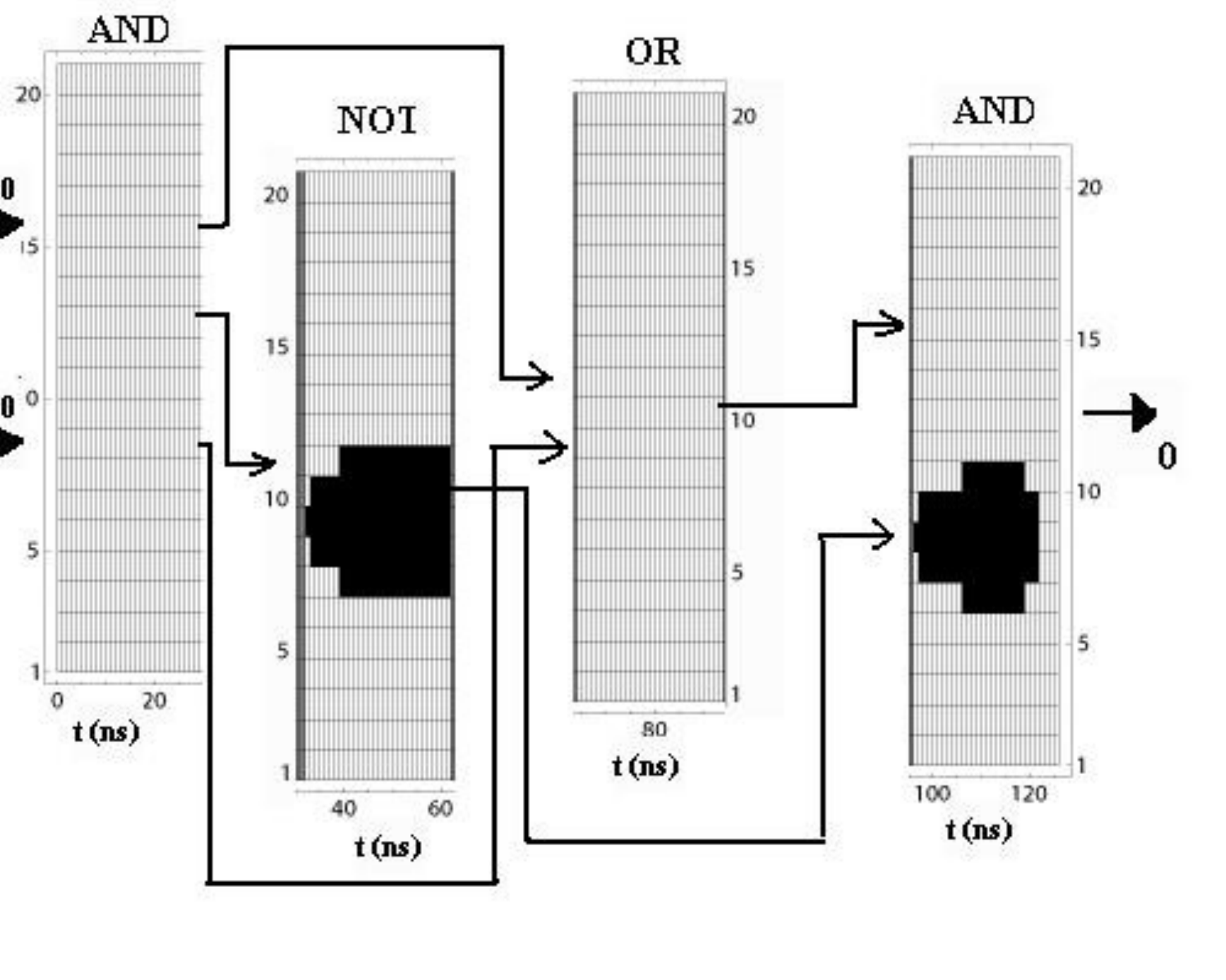}}
		\subfigure[]{\includegraphics[width=0.49\linewidth]{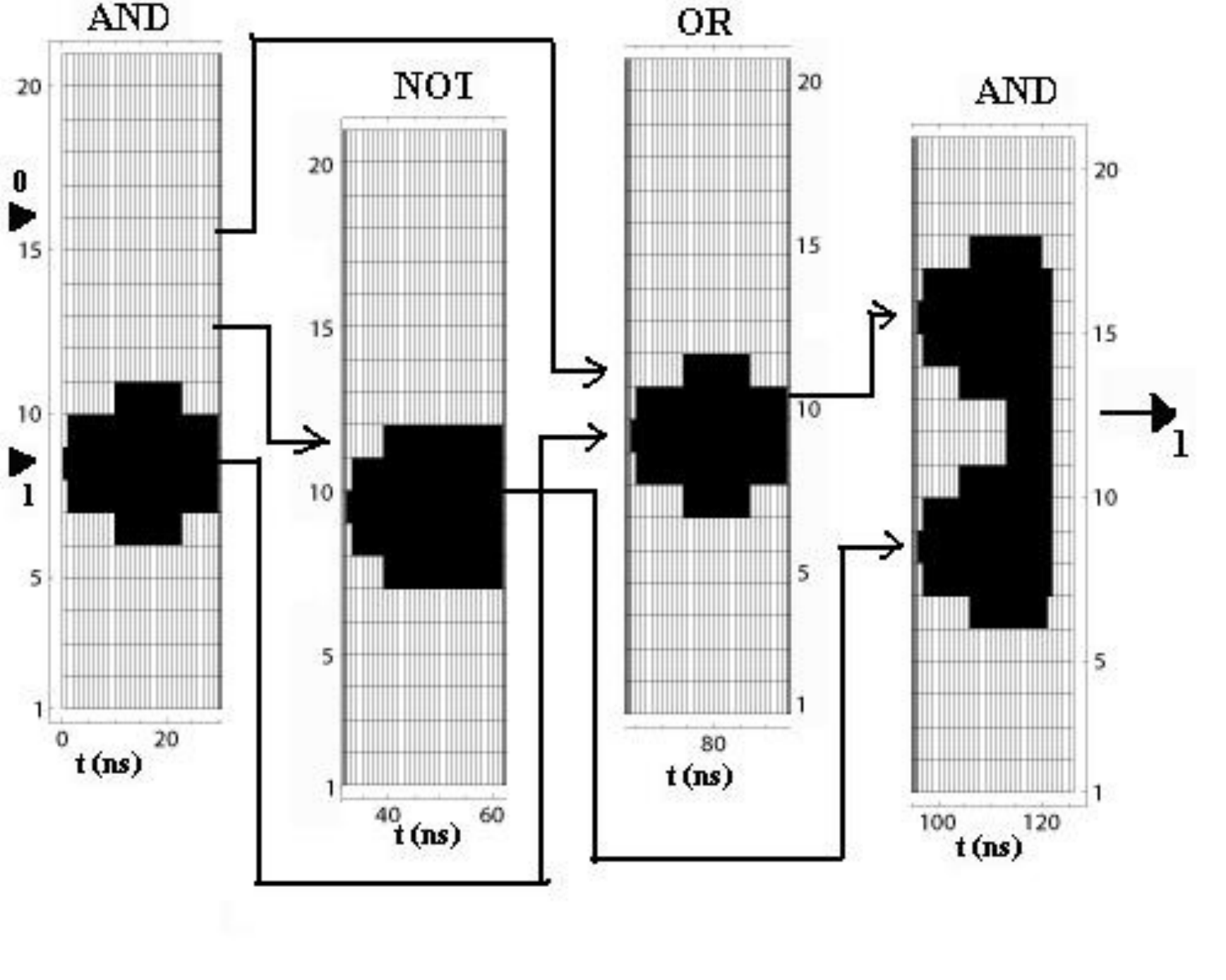}}
		\subfigure[]{\includegraphics[width=0.49\linewidth]{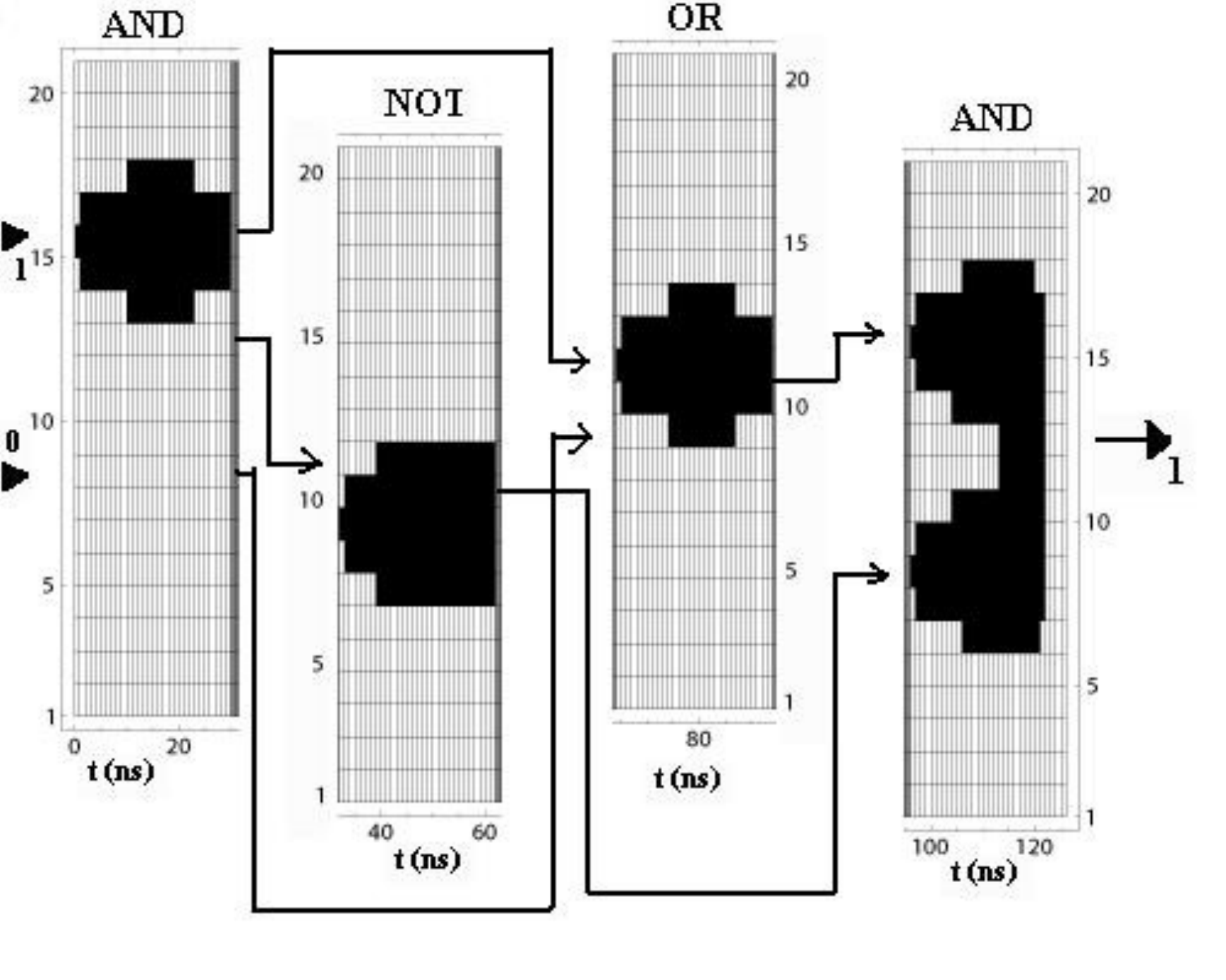}}
		\subfigure[]{\includegraphics[width=0.49\linewidth]{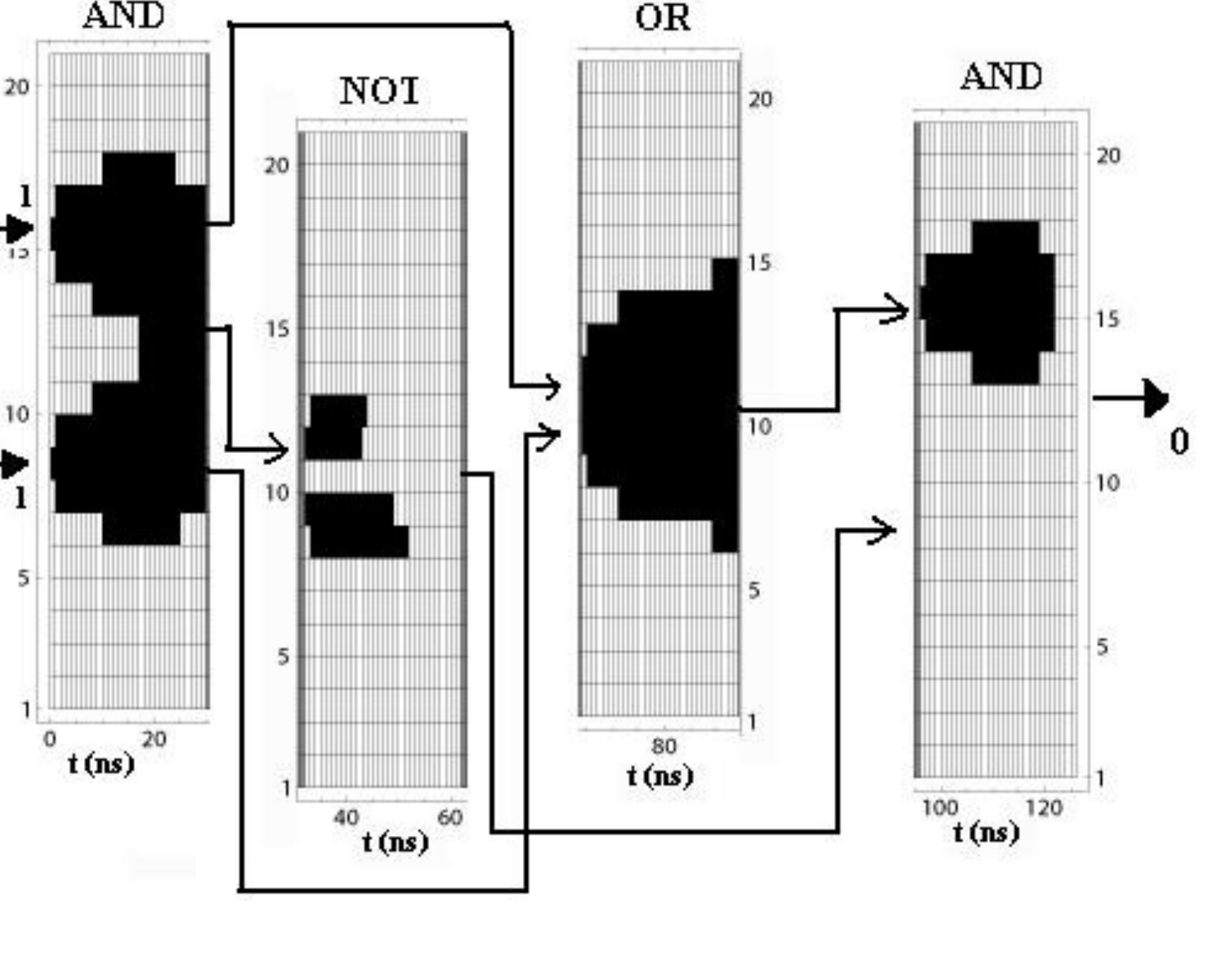}}
		\caption{Unforced pulses, {\sc xor} operation. Inputs are (a)~00, (b)~10; (c)~01 and (d)~11 in cells 8 and 15 of the leftmost
		gate; output in cells 11 and 12 of the rightmost. Cells excited above $0.1 V_0$ are black; time evolves along the horizontal axis and cells are aligned along the vertical axis.}
		\label{xor_tuz}
\end{figure}

To build an {\sc xor} gate we use the formula 
\begin{equation}
a \oplus b = (\neg (a \wedge b)) \wedge (a \vee b) 
\end{equation}
as shown in Fig.~\ref{xor_tuz}.
Each run is composed by four gates, and it is supposed that some mechanism  reads the output of each gate,
amplifies the signal and passes the signal to the next gate. In the figure the arrays represent signals
travelling between gates.
The first from the left is an {\sc and} gate with input cells 8, 15 and output cell 12;
its output goes to the next gate: the {\sc not} gate having input in cell 11 and output in cell 10. The original inputs,
coming out at cells 8 and 15 of the first gate, are sent to input cells 9 and 11 of the {\sc or} gate in the third section. Finally, the output of the {\sc or} gate in cell 10 and the output of the {\sc not} gate are sent to input cells 8 and 15 of the final {\sc and} gate. The final output of the {\sc xor} operation is found in cells 11 and 12 at the right end.


\section{Gates with forced pulses}

Let us we consider interactions and possible gates implemented by applying input pulses continuously at input sites. Namely, we send a sinusoidal pulse to input during the whole evolution of the system, with the intensity of the input pulse remaining constant.

We use slightly different values for resistances: 
$R_1=9.23 \, 10^{6} \Omega, R_2=1.32 \, 10^{6} \Omega$, that can be obtained
changing concentrations of $K^{+}$ and $Na^{+}$;
excitations are sinusoidal with a 1~ns period.

\begin{figure}[!tbp]
	\centering
		\subfigure[]{\includegraphics[width=0.49\linewidth]{and00_tuz}}
		\subfigure[]{\includegraphics[width=0.49\linewidth]{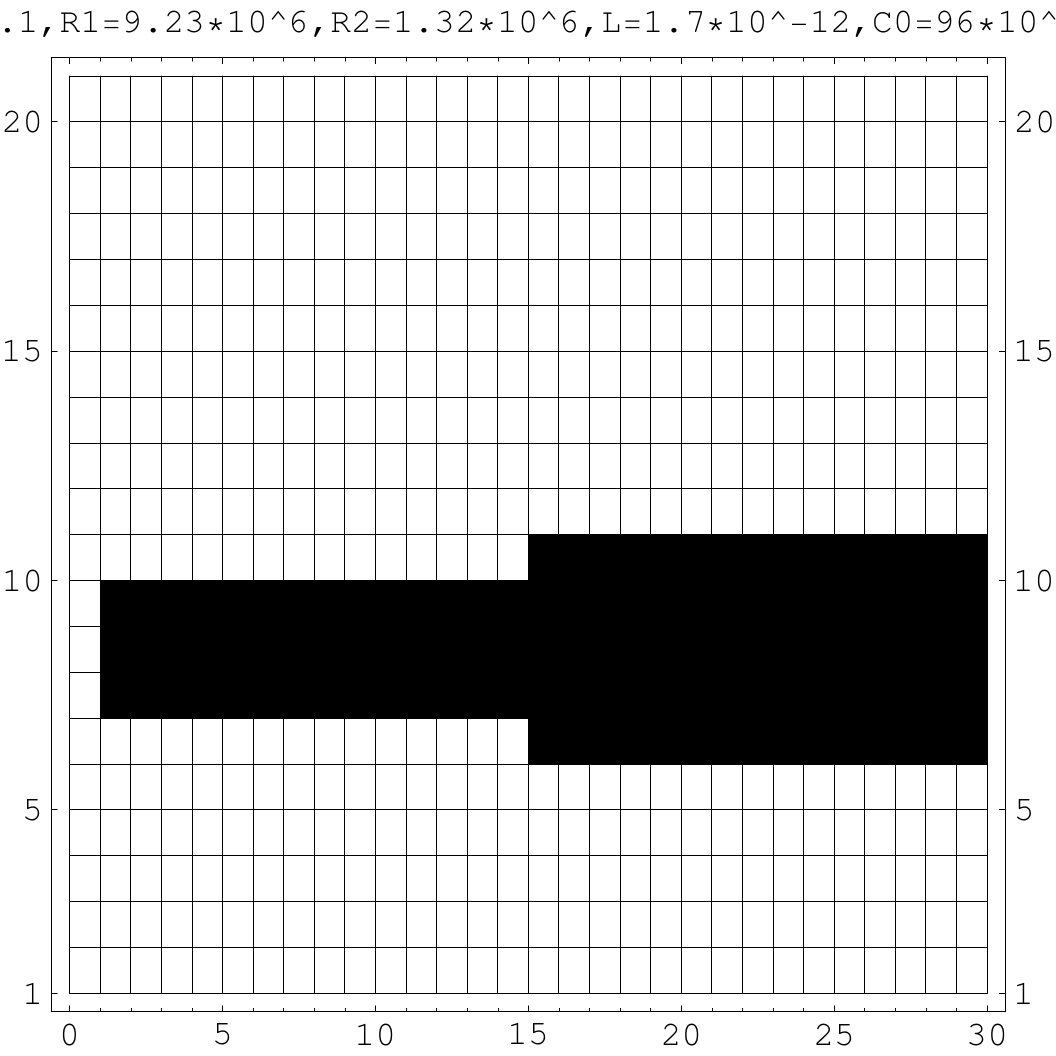}}
		\subfigure[]{\includegraphics[width=0.49\linewidth]{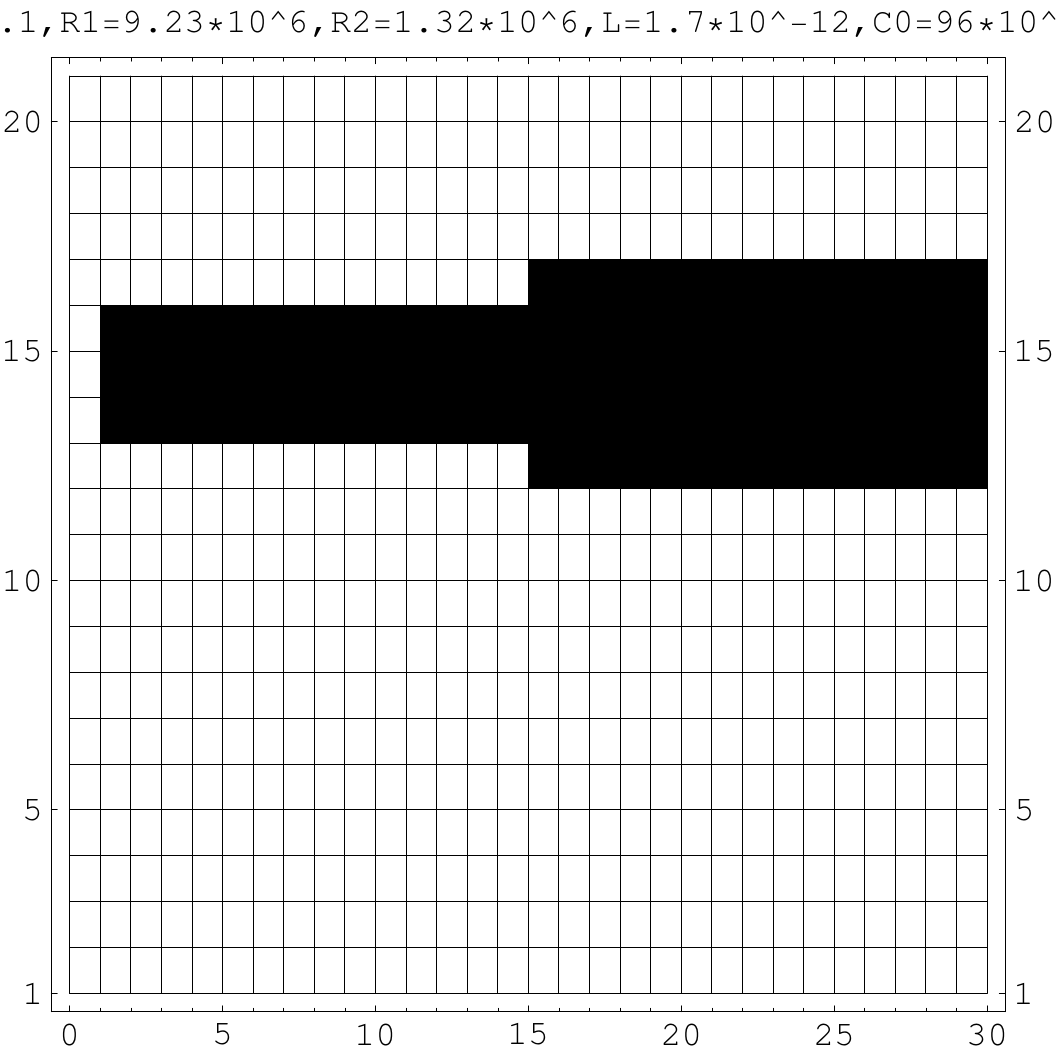}}
		\subfigure[]{\includegraphics[width=0.49\linewidth]{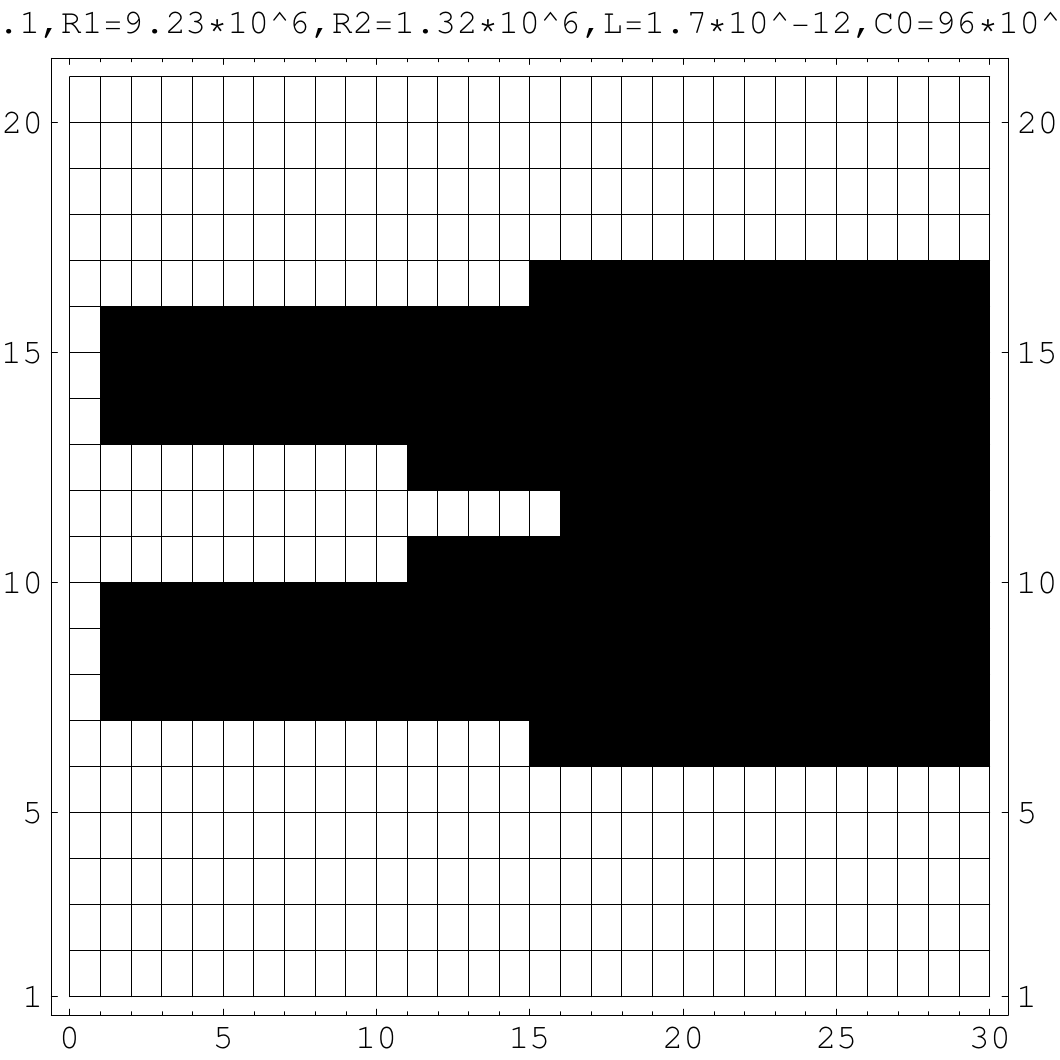}}
		\caption{Forced pulses, {\sc and} operation. Inputs are (a)~00, (b)~10, (c)~01 and (d)~11 in cells 8 and 14; output in cell 11.}
		\label{and_tuzfor}
\end{figure}

Figure~\ref{and_tuzfor} shows the {\sc and} operation obtained with two input cells at positions 8 and 14. The output is found in cells 11 that is excited at the end of the evolution if both the input cells are excited.

\begin{figure}[!tbp]
	\centering
	\subfigure[]{\includegraphics[width=0.49\linewidth]{and00_tuz}}
	\subfigure[]{\includegraphics[width=0.49\linewidth]{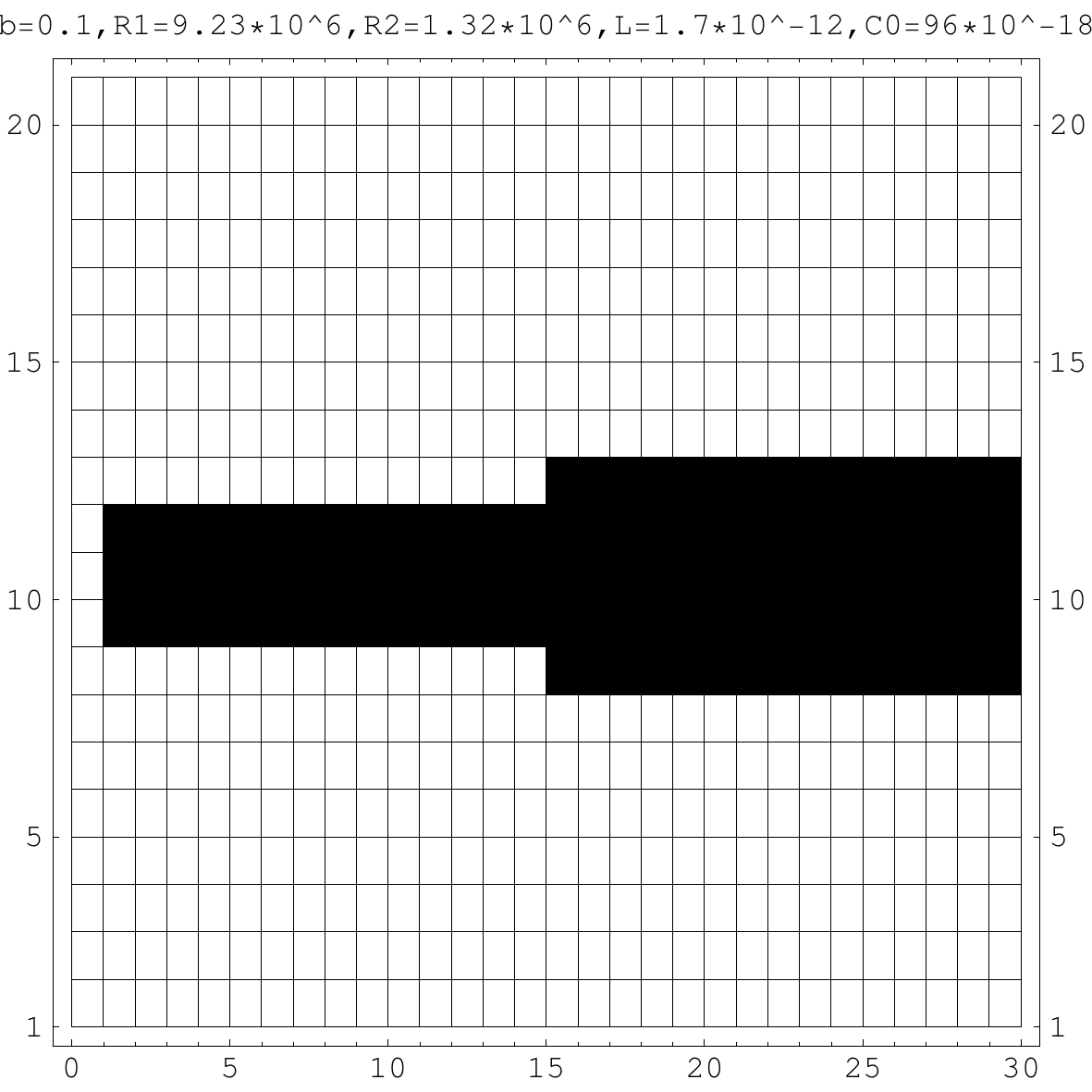}}
		\subfigure[]{\includegraphics[width=0.49\linewidth]{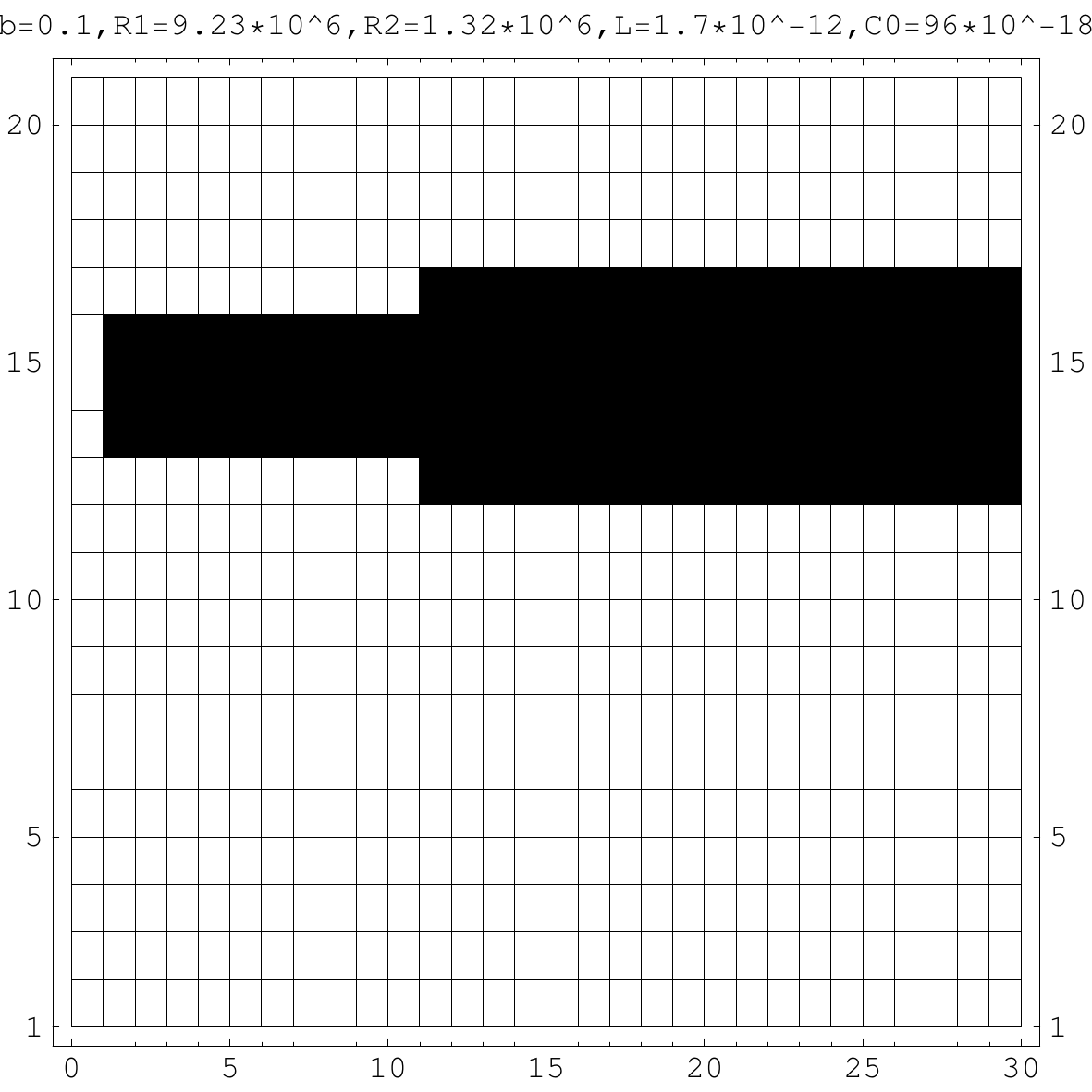}}
		\subfigure[]{\includegraphics[width=0.49\linewidth]{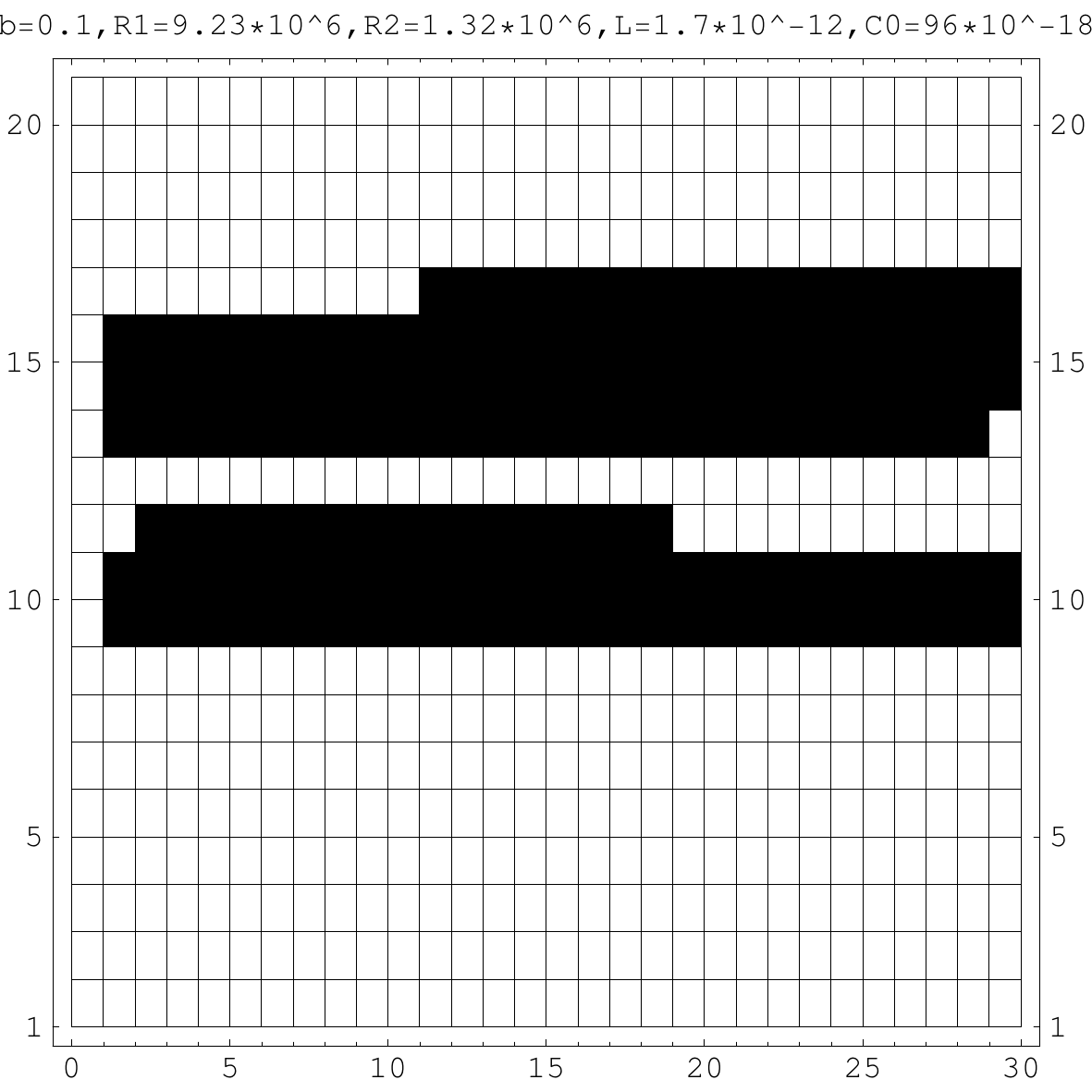}}
		\caption{Forced pulses, {\sc xor} operation. Inputs are (a)~00, (b)~10, (c)~01 and (d)~11 in cells 10 and 14; output in cell 12.}
		\label{xor_tuzfor}
\end{figure}

Figure~\ref{xor_tuzfor} shows the {\sc xor} operation obtained with two input cells at positions 10 and 14. The output is found in cells 12 that is excited at the end of the evolution if only one the input
cells is. The input pulses are sinusoidal with opposite phases.
The other logical gates can be implemented in a similar way.

\begin{figure}[!tbp]
	\centering
	\subfigure[]{\includegraphics[width=0.49\linewidth]{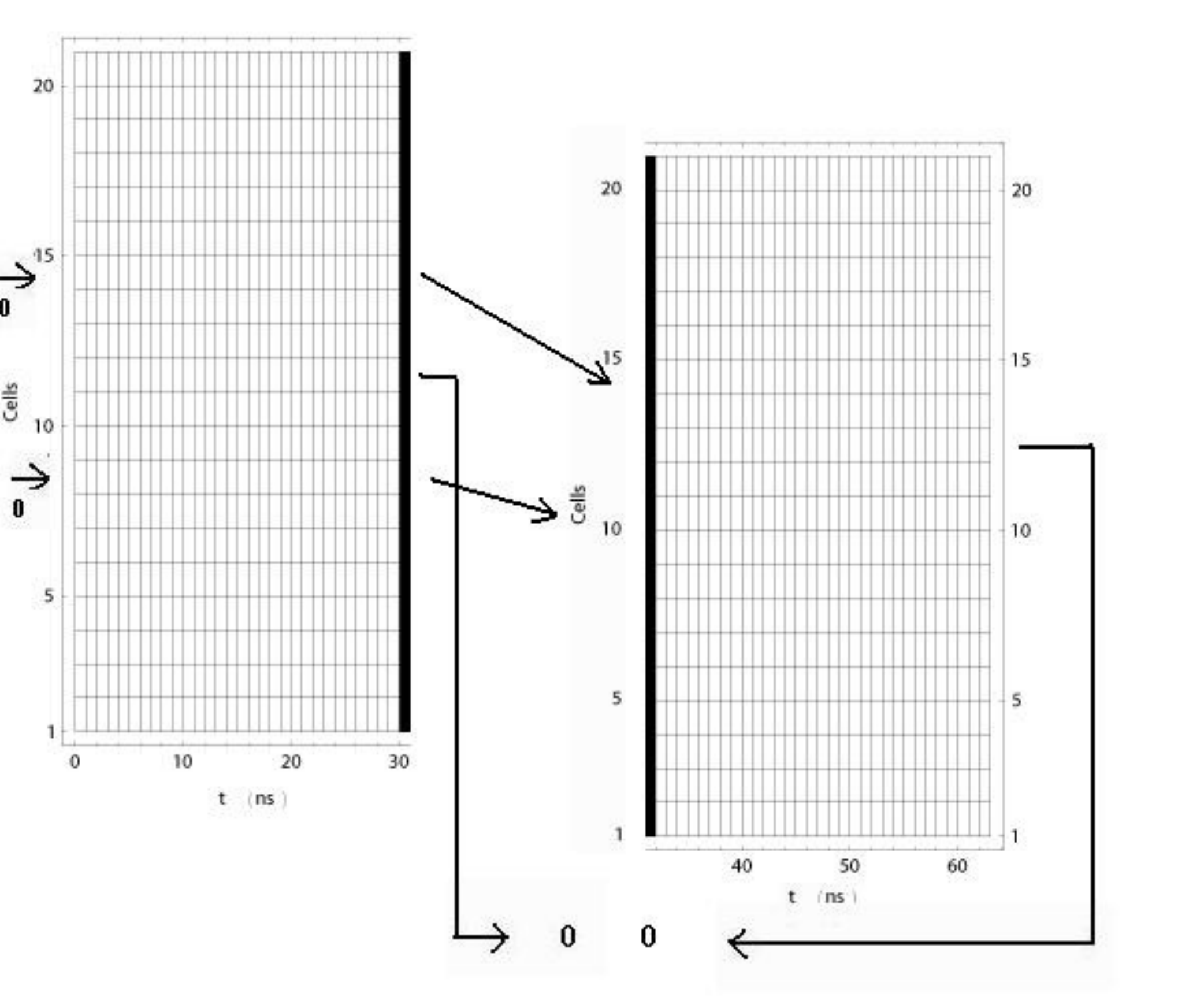}}
	\subfigure[]{\includegraphics[width=0.49\linewidth]{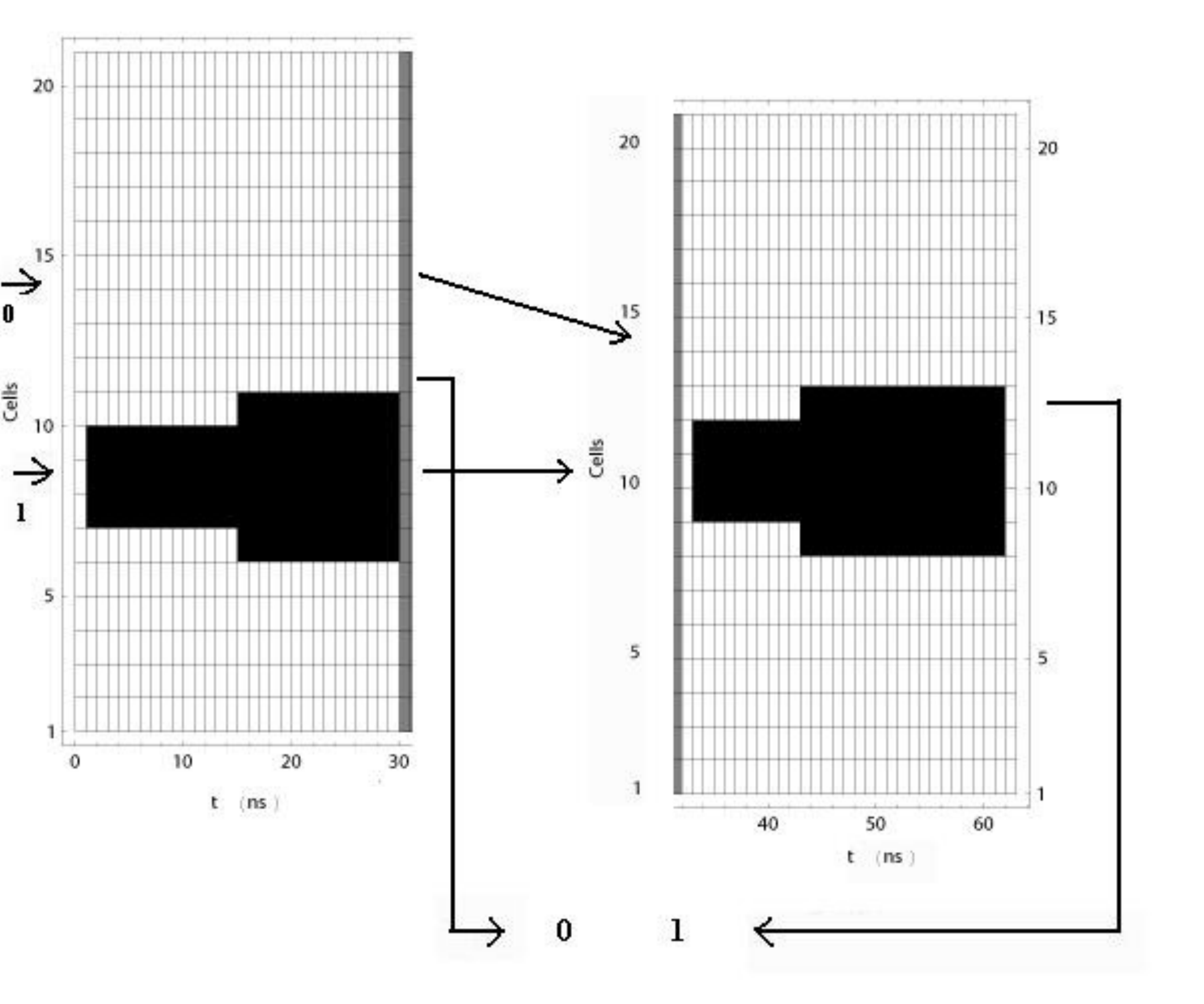}}
		\subfigure[]{\includegraphics[width=0.49\linewidth]{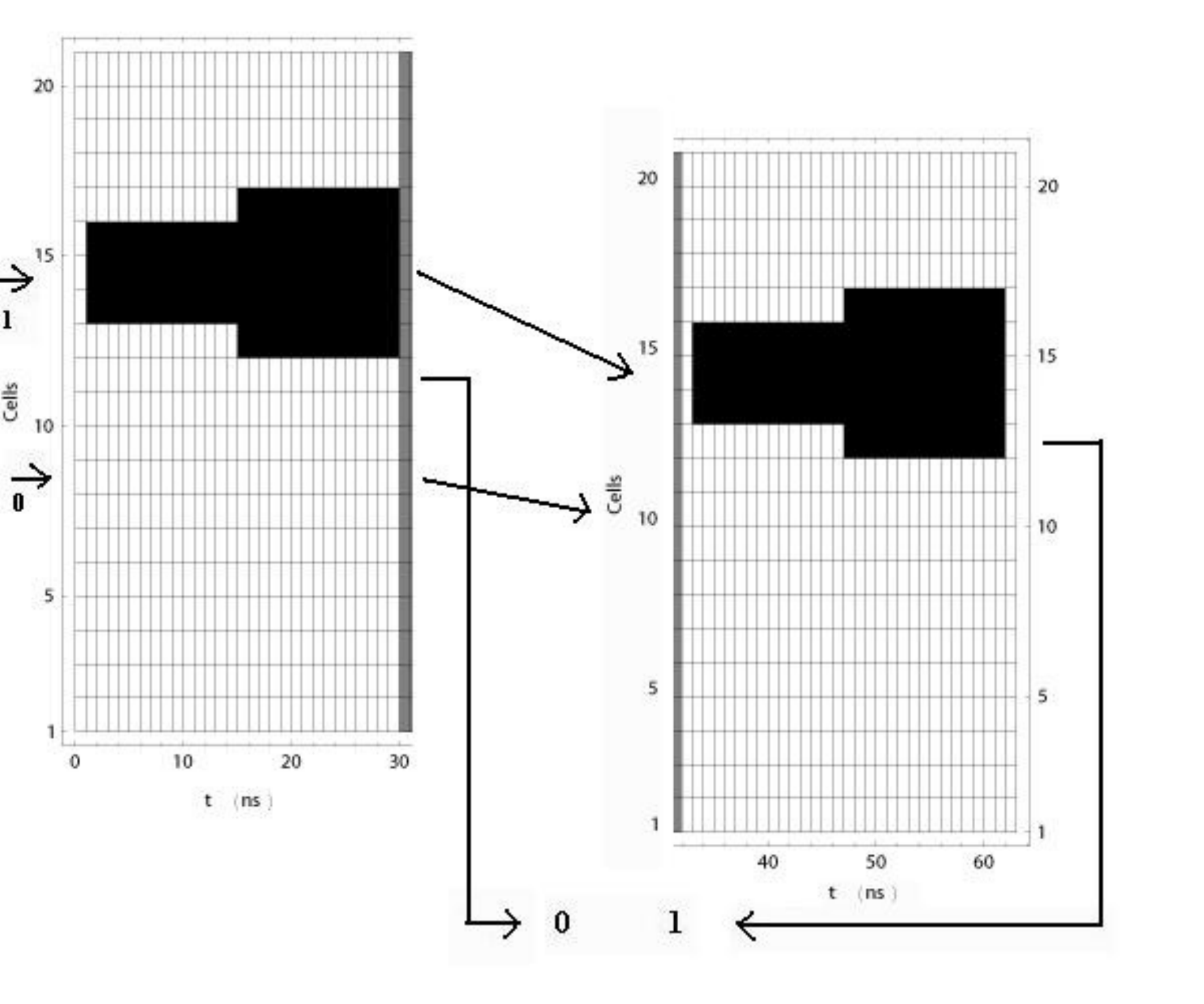}}
		\subfigure[]{\includegraphics[width=0.49\linewidth]{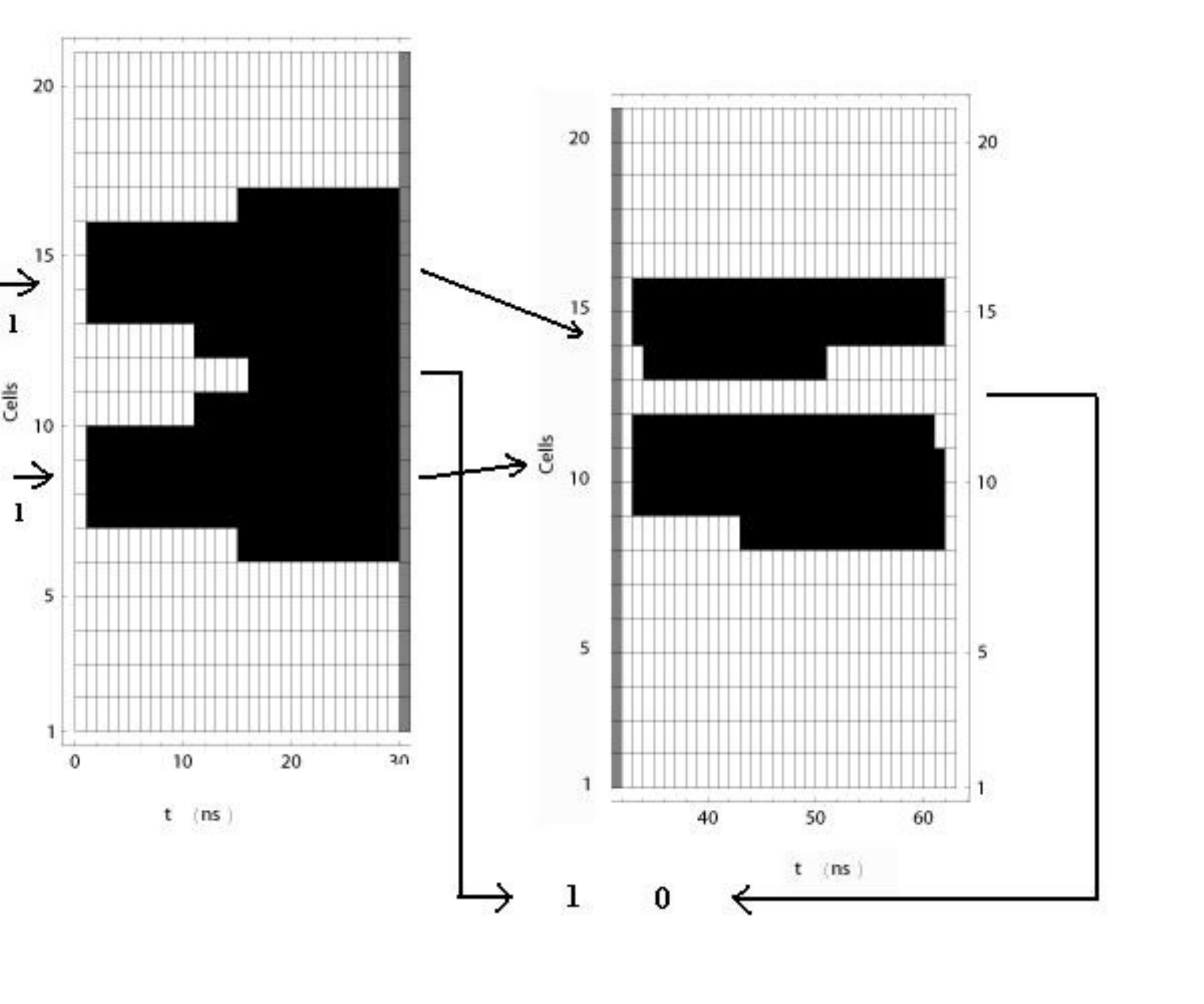}}
		\caption{Forced pulses, half-adder. Inputs are (a)~00, (b)~10, (c)~01 and (d)~11 in cells 8, 10 and 14; outputs in cell 11 and 12.}
		\label{hadd_tuzfor}
\end{figure}

We also show in Fig.~\ref{hadd_tuzfor} a one-bit half-adder: it is composed by an {\sc and} gate followed by an {\sc xor}. The {\sc and} gate has
input cells 8 and 14 and outputs the most significant bit (the carrier) in cell 11; 
the {\sc xor} has input cells 10 and 14 and outputs the less significant bit in cell 12. It is supposed that input in cell 14
is common to both gates, while input in cell 8 of the AND gate travels to cell 10 of the {\sc xor} changing its phase.

One could argue that it could be difficult in practice to send an input pulse to exactly one monomer
(or to initialize it in an
excited status), which requires a localization precision of under 5~nm. We therefore compute some evolutions supposing that the pulse 
hits a number $N$ of monomers. Under these conditions, the values of resistance, inductance and capacitance 
must be computed with appropriate addition rules. For resistance we have:
\begin{equation}
R_{2\,tot} = ( \sum_{i=1}^{N} R_{2,i}^{-1} )^{-1}, \; R_{1\,tot} =  \sum_{i=1}^{N} R_{1,i}
\label{sumres}
\end{equation}

\begin{figure}[!htb]
	\centering
		\subfigure[]{\includegraphics[width=0.48\linewidth]{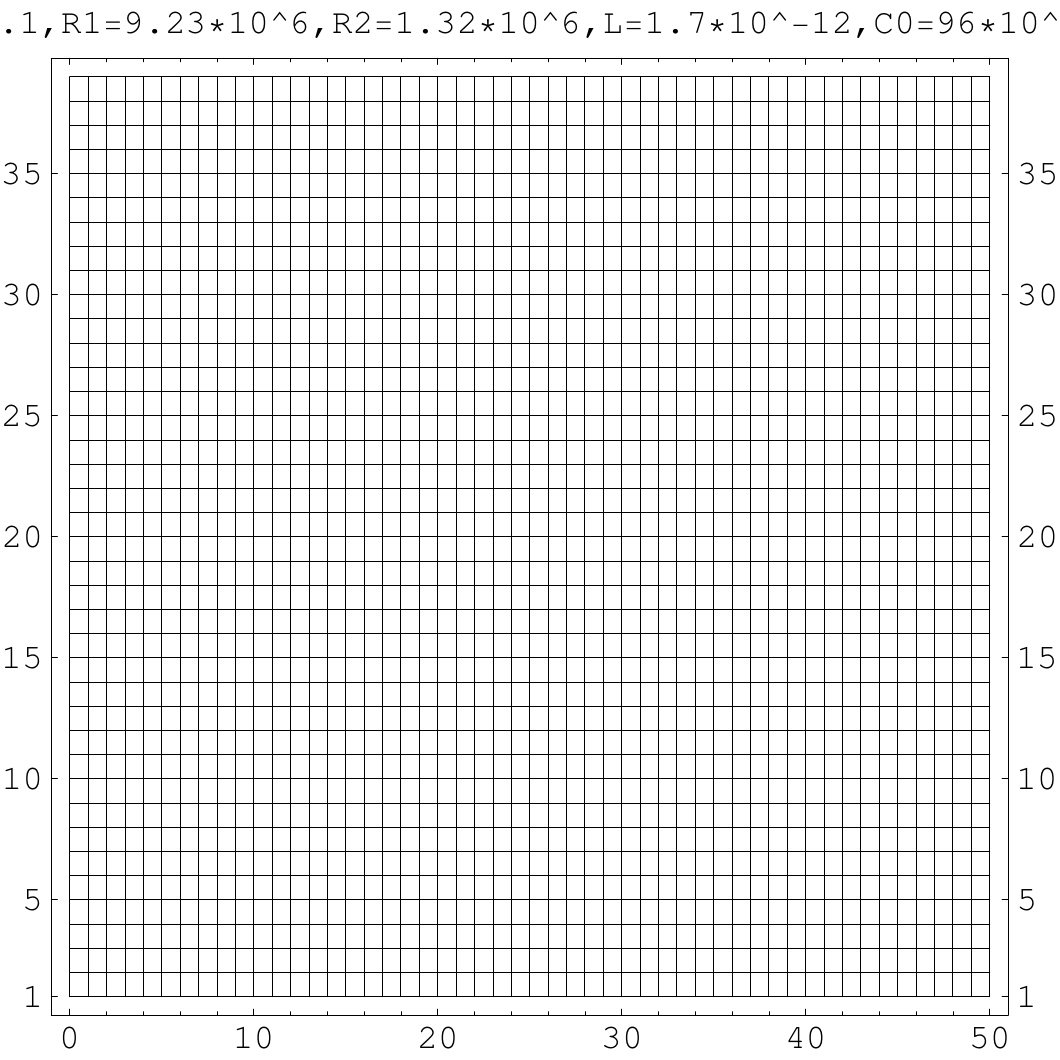}}
		\subfigure[]{\includegraphics[width=0.48\linewidth]{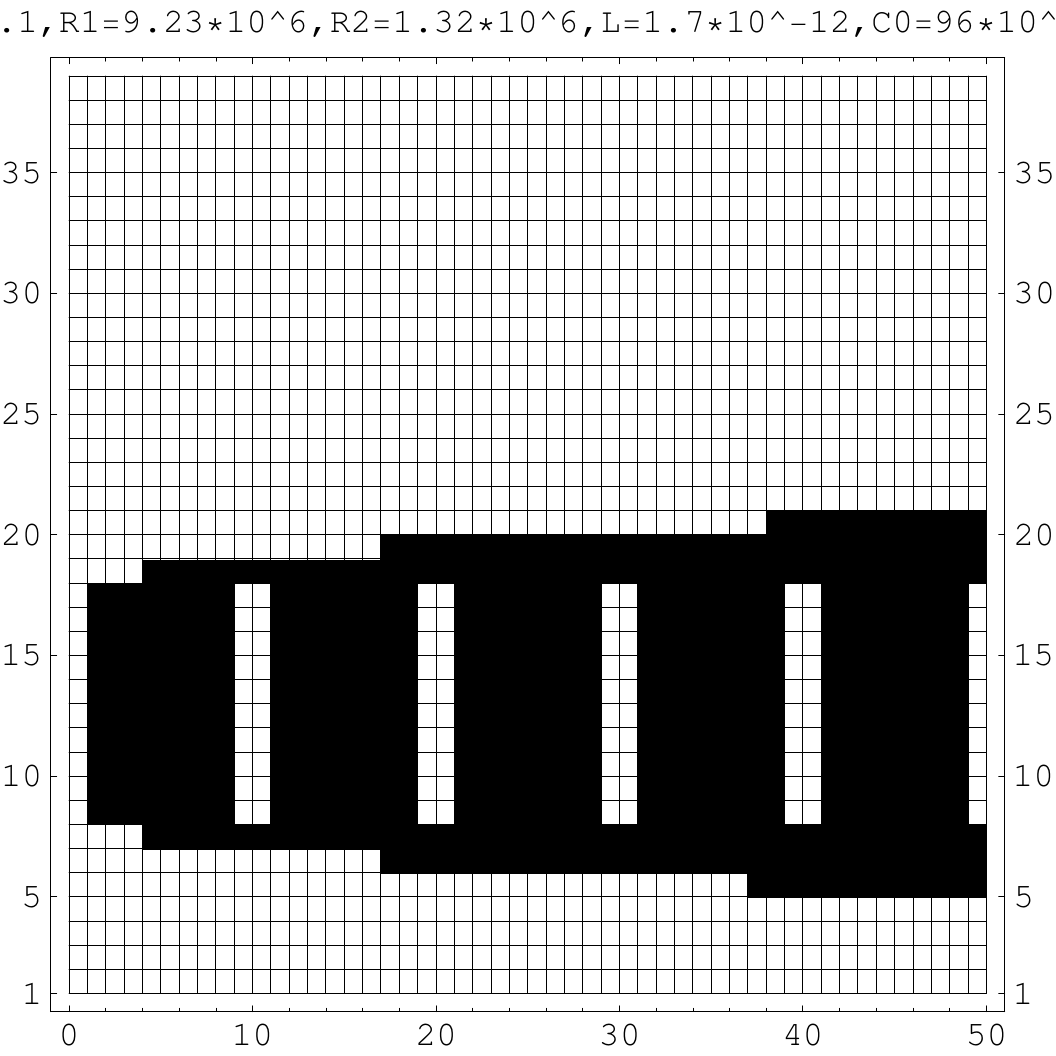}}
		\subfigure[]{\includegraphics[width=0.48\linewidth]{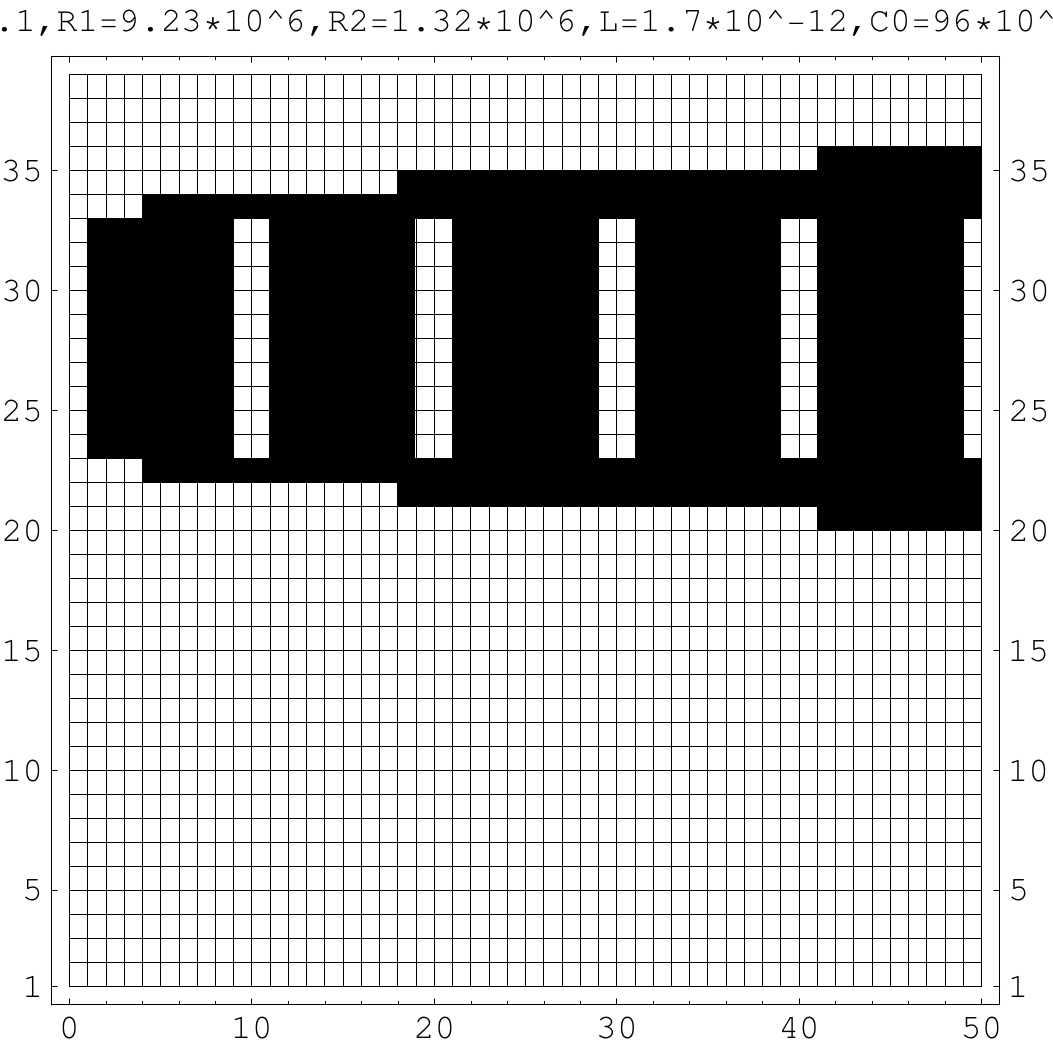}}
	\subfigure[]{\includegraphics[width=0.48\linewidth]{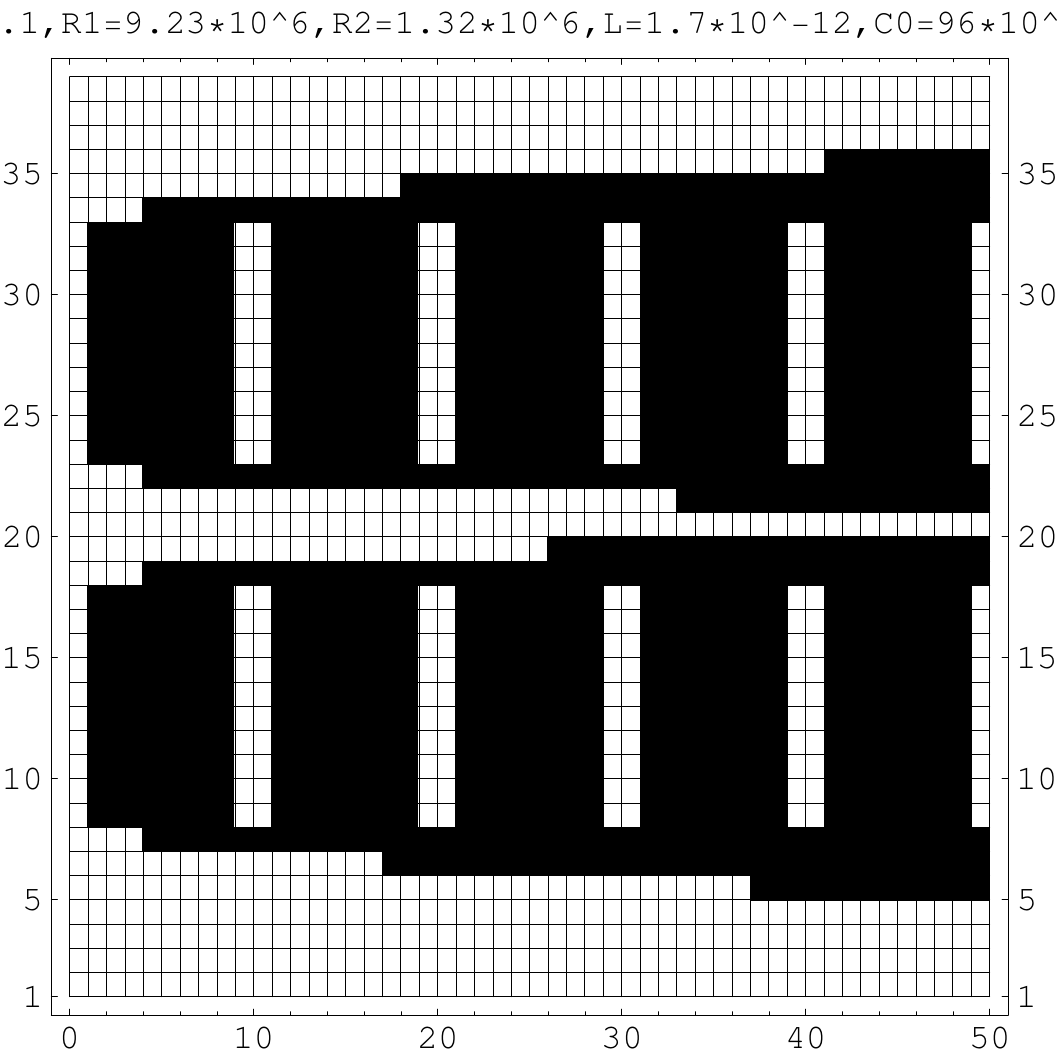}}
		\caption{Forced pulses, {\sc xor} operation. Inputs are (a)~00, (b)~10, (c)~01 and (d)~11 in cells 8-17 and 23-32; output in cell 20.}
		\label{xor_tuzfor10}
\end{figure}

Total inductance and capacitance are obtained summing those of individual monomers.
Figure~\ref{xor_tuzfor10} shows the {\sc xor} gate with input sent to groups of 10 monomers in positions 8-17 and and 23-32. As in Fig.~\ref{xor_tuzfor}, the input pulses are sinusoidal with opposite phases.
 Output is in cell 20. 
In our approximation, we do not consider the detailed evolution inside the groups of monomers receiving the input, they are treated as a single cell, with $R$, $L$, $C$ parameters computed summing on 10 monomers.

We note that the results are  robust: pulses fade slowly and a threshold of 0.3 of the input signal can be used, instead of 0.1 as in the previous computations.

\section{Conclusion}

In numerical experiments presented in this paper we have shown that it is possible to build basic logical gates from actin filaments considered as RLC circuits, provided that input pulses can be sent to specific locations and initial values of voltage can be 
set in actin monomers. We demonstrated that several logical gates can be implemented via interactions between input pulses. For example, to realise an {\sc and} gate we send the same phase pulses to the input cells. The pulses sum with each other. The output threshold is set in such manner that if there is only one active input the output potential is lower than the threshold. If there are two summed input potentials the threshold is exceeded. To realise an {\sc xor} gate we send pulses with opposite phases so when they meet they are subtracted from each other. 

To try a real implementation it is necessary to specify how output can be measured with sufficient
sensitivity (either 0.1 or 0.3 of the initial voltage) and how output pulses of a gate can be reliably sent
as input to another. These points will be considered in future works, we can however anticipate some considerations 
about cascading gates.


Actin filaments are usually found in networks, forming cross links that can have a parallel or orthogonal shape. 
These links can be made by several kinds of binding proteins that originate different network topologies.
We consider for instance filamin A (see e.g. \cite{nakamura2007structural}), that builds orthogonal networks: where two actin filaments cross link, we suppose
that electric pulses generated as output of a gate in a filament can reach the other using the filamin molecule as
a connecting wire. Branching of actin filaments networks has been observed in cells and it requires the presence of ARP2/3 protein as a branch point from which two actin filaments propagate at acute angles~\cite{higgs2001regulation}.

Considering for the sake of simplicity the same electrical mechanisms as for actin, we could compute $R_1$, $R_2$, $C_0$, 
and $L$ for filamin, taking into account just its dimensions. 

In this way, in principle, any number of actin gates could be connected, however there is a problem:
pulses intensity will fade away rather quickly, unless some amplifying method is found. 

Some studies (e.g. \cite{sampson2003direct}) have
found additional roles of filamin A in signal transduction, and others like \cite{schubert2004actin} have found that actin filaments can regulate some voltage-gates ion channels. We can therefore think of a system where the output voltage of a gate opens a ion channel that in turn activates the next gate and so on. Of course this subject must be treated in more depth to
draw any practical conclusions, as the spatial distribution of the ion channels within the actin network, their dimensions,
the necessary electrochemical gradients, the type of ions that
can be used, their diffusion rates etc. must be carefully evaluated to check the feasibility of a logical device.

Tubulin is another important structural protein that polymerises into cylindrical microtubule structures that have been found to be electrical amplifiers~\cite{priel2006biopolymer}. Actin filaments are connected to tubuline microtubules via microtubule associate proteins (MAP) ~\cite{griffith1978evidence,dehmelt2004actin}. Thus, if electrical properties of the MAP are similar to that of actin units we can speculate that cascading of Boolean gates into  arithmetic circuits can be done via actin to MAP to  tubuline to MAP to actin links. It has been recently proposed that microtubule phosphorylation by calmodulin kinase II enzyme complex can be linked to memory formation at a cellular level~\cite{craddock2012cytoskeletal}. In a manner similar to that presented in this paper for actin filaments, signals propagating along microtubules have been also shown to be capable of interacting via logical gates~\cite{craddock2012cytoskeletal}. This can lead to an integration of signal storage and processing capabilities of the cytoskeleton, which is composed of both actin filaments and microtubules that interact with each other.

In case of \emph{in vitro} desing of actin-based computing circuits a tailored mono-molecular amplifiers~\cite{joachim2000electronics} can be inserted between actin strands in cascaded gates.

\bibliography{actin_rlc}

\end{document}